\documentclass[]{aastex631}
\usepackage{amsmath}
\usepackage{amssymb}

\shorttitle{Crystalline ratio to amorphous in comets}
\shortauthors{Okamoto \& Ida}
\graphicspath{{./}{figures/}}
\begin{document}

\title{Monte Carlo Simulation of Dust Particles in a Protoplanetary Disk: Crystalline to Amorphous Silicate Ratio in Comets}
\author[0000-0003-1545-2723]{Tamami Okamoto}
\affiliation{Earth-Life Science Institute \\
Tokyo Institute of Technology \\
Meguro-ku, 152-8550 Tokyo, Japan}

\author{Shigeru Ida}
\affiliation{Earth-Life Science Institute \\
Tokyo Institute of Technology \\
Meguro-ku, 152-8550 Tokyo, Japan}

\begin{abstract}
Observationally inferred crystalline abundance in silicates in comets,
which should have been formed in the outer region of a protoplanetary disk,
is relatively high ($\sim 10$--60\%), although crystalline silicates would be formed by annealing of amorphous precursors in the disk inner region.
In order to quantitatively address this puzzle, 
we have performed Monte Carlo simulation of advection/diffusion of silicate particles in a turbulent
disk, in the setting based on pebble accretion model:
pebbles consisting of many small amorphous silicates embedded in icy mantle are formed in the disk outer region,
silicate particles are released at the snow line, crystalline silicate particles are produced
at the annealing line, the silicate particles diffused beyond the snow line,
and they eventually stick to drifting pebbles to come back to the snow line.
In a simple case without the sticking and with a steady pebble flux,
we show through the simulations and analytical arguments 
that crystalline components in silicate materials beyond the snow line
is robustly and uniformly $\simeq 5$\%.
On the other hand, in a more realistic case 
with the sticking and with a decaying pebble flux, the crystalline abundance is raised up to $\sim 20$--25\%,
depending on the ratio of decay and diffusion timescales. 
This abundance is consistent with the observations. 
In this investigation, we assume a simple steady accretion disk. 
The simulations coupled with the disk evolution 
is needed for more detailed comparison with observed data.

\end{abstract}
\keywords{protoplanetary disk --- comets --- silicate}

\section{Introduction} \label{sec:intro}
Crystalline silicates, such as forsterite, are formed by 
re-condensation after sublimation at $T \ga 1400\,{\rm K}$  
of amorphous precursors
or annealing of them at $T \ga 1000\,{\rm K}$ 
\citep[e.g.,][]{Wooden2008}. 
However, they are observed in 
outer cold regions of protoplanetary disks and the {\it Stardust} samples
\citep[e.g.,][and references therein]{Brownlee2006, Ciesla2011}.
Infrared observations of comets also show
10--60\% of silicates are crystalline \citep[e.g.,][also see compiled data for several comets in Shinnaka et al. 2018]{Sitko2011, Harker2011}.
Comets were formed in outer region of a protoplanetary disk, such as Kuiper belt or Uranus-Neptune zone, and scattered to Oort Clouds \citep[e.g.,][]{Dones2015}. 
Some mechanism should have transported the crystalline silicate particles
that experienced $T \ga 1000\,{\rm K}$ in the inner disk region to the outer region.
The radial turbulent diffusion is one of available transporting mechanisms \citep[e.g.,][]{Gail2001,Bockelee2002,Cuzzi2003},
while centrifugal jets \citep[][]{Shu1996} or global disk flow \citep[e.g.,][]{Keller2004,Ciesla2007,Desch2007} could also be available.
In this paper, we focus on the radial turbulent diffusion.

\citet{Ciesla2010,Ciesla2011} developed a 3D Monte Carlo simulation code 
for turbulent diffusion of crystalline silicates in a protoplanetary disk
to successfully show that the particles initially at $\sim 5$ au can
diffuse out beyond 20 au.
However, because the initial distributions of amorphous and crystalline silicates
were not clear, the results were not compared quantitatively with the
observed abundance of crystalline components in silicates in the comets.

\citet{Pavlyuchenkov2007} numerically solved one-dimensional diffusion equation
for crystalline materials formed at the formation front in the inner disk region.
\citet{Arakawa2021} discussed the modulations from the
\citet{Pavlyuchenkov2007}'s results by silicate particle growth and by drift due to gas drag
in the disks with and without a pressure bump created by magnetically driven disk winds.
Both papers assumed that amorphous silicates are stationarily distributed 
with the uniform solid-to-gas ratio in the steady accretion disk,
and \citet{Arakawa2021} concluded that the crystalline abundance is very low 
unless a global pressure bump is introduced.
However, their setting on silicate particles is not consistent with
the new model of planetary accretion called ``pebble accretion" \citep[e.g.,][]{Ormel2010,Lambrechts2014}.

In the pebble accretion model,
pebbles with cm--m size are formed in disk outer regions, drift
all through the disk, and some fraction of the drifting pebbles are accreted by planetary seeds \citep[e.g.,][]{Lambrechts2014,Sato2016}.    
The pebbles would consist of many small amorphous silicate particles 
embedded in icy mantle and the silicate particles are released
by ice sublimation at the snow line \citep[e.g.,][]{Saito2011,Morbidelli2016,Ida&Guillot2016}.
The reason to consider this pebble structure is that
 the threshold velocity for fragmentation of silicates 
would be generally one order smaller than that of ices \citep[e.g.,][]{Blum2000,Zsom2011,Wada2011},
although this conventional view is recently challenged \citep[][]{Kimura2015,Gundlach2018,Musiolik2019,Steinpliz2019}.
Assuming this silicate-ice structure of pebbles, 
\cite{Ida2021} derived the conditions for the gravitational instability
of the pile-up of the released silicate particles by local 2D ($r-z$) 
Monte Carlo simulation, following the method by \citet{Ciesla2011}. 

The motivation of this paper is to apply the above setting of
injection of amorphous silicate particles at the snow line for predicting 
the crystalline silicate ratio in the total (silicate and amorphous) silicates in disk outer regions.
We find that about 5\% of the released amorphous particles drifting inward from the snow line
come back to disk outer regions by outward diffusion, 
after experiencing high temperature $T \ga 1000\,{\rm K}$ to be 
transferred to crystalline particles. 
To calculate the global evolution of silicate dust particle distributions, we adopt
global 3D (\(x-y-z\)) Monte Carlo simulation by \cite{Ciesla2011}. 
We do not apparently incorporate a coupled silicate
dust growth such as the model by \cite{Misener2019}, because the results here do not change,
as long as Stokes number of the silicate particles is smaller than the \(\alpha\) parameter of the turbulent
mixing, while the effect of the silicate growth is discussed in Appendix.
On the other hand, we include the effect of rapid inward drift of the silicate particles that coagulate
with drifting pebbles, which was discussed by \cite{Misener2019}.
Because we can calculate surface densities both of
crystalline and amorphous silicate particles as a function of orbital radius,
a quantitative comparison between theoretical prediction and the observation of comets
can be done.

In Section \ref{sec:method}, we describe the protoplanetary disk model that we adopt
and Monte Carlo methods for advection due to gas drag and turbulent diffusion of silicate particles in the disk. The sticking rate of the particles to drifting pebbles beyond the snow line is also evaluated.
In Section \ref{sec:analy}, we describe the analytical steady solution of the radial advection-diffusion equation to explain the simulation results with and without silicates sticking to pebbles. In Section \ref{sec:resu}, we show the simulation results 
for the simple case neglecting the sticking (Section \ref{subsec:nopeb}), the case with sticking to icy pebbles (Section \ref{subsec:peb}), and the case with the attenuation of a pebble flux due to the depletion of solid materials in the reservoir in the outer disk
(Section \ref{subsec:decay}). 
While the sticking to the pebbles inhibits outward diffusion of the silicate particles, 
the decaying pebble flux tends to diminish the sticking effect.
We show that the radial diffusion in the case with the decaying pebble flux can 
quantitatively explain the relatively high abundance in the comets.
Section 5 is the conclusion and discussion.

\section{Method} \label{sec:method}
\subsection{Disk model}
\label{subsec:disk}

We adopt the quasi-steady composite disk model derived by \cite{Ida2016} and \cite{IdaYmaOku2019}.
For simplicity, the depletion due to photoevaporation is neglected
and we assume that the disk evolution is regulated by turbulent viscous diffusion.
In general, the viscous heating is dominant in the inner disk regions,
while the irradiative heating dominates in the outer disk regions.
We composite these two regions by the disk mid-plane temperature 
$T = \max(T_{\rm vis}, T_{\rm irr})$,
where \(T_{\textrm{vis}}\) and \(T_{\textrm{irr}}\) 
are the temperature in the viscous-heating 
and irradiation dominated regions given respectively by 
\begin{align}
    \label{Tvis}
    T_{\textrm{vis}} & \simeq 330\left(\frac{M_*}{1.0M_{\odot}}\right)^{3/10}\left(\frac{\alpha}{10^{-2}}\right)^{-1/5}\left(\frac{\dot{M}_\textrm{g}}{10^{-7}M_{\odot}/\textrm{yr}}\right)^{2/5}\left(\frac{r}{1\,\textrm{au}}\right)^{-9/10} \textrm{K}, \\
    \label{Tirr}
    T_{\textrm{irr}} & \simeq130\left(\frac{L_*}{1.0L_{\odot}}\right)^{2/7}\left(\frac{M_*}{1.0M_{\odot}}\right)^{-1/7}\left(\frac{r}{1\,\textrm{au}}\right)^{-3/7} \textrm{K},
\end{align}
and where \(\dot{M}_\textrm{g}\) is the gas accretion rate through the disk and \(\alpha\) is a parameter of turbulent viscosity,
which are scaled by the fiducial parameters in our disk model,
$\dot{M}_\textrm{g}=10^{-7}M_{\odot}/\textrm{yr}$ and $\alpha=10^{-2}$.

The temperature in the viscous-heating depends on the opacity.
Equation~(\ref{Tvis}) corresponds to relatively low opacity with relatively large silicate particles
with the size $\sim 0.1\, \rm mm$ and the temperature is higher for $\mu$m size particles \citep{Oka2011}.
As we will show in section 4.1, as long as both the snow and the annealing lines are
in the viscous-heating dominated regime,
the predicted crystalline silicate abundance is not affected by the numerical factor of $T$;
the abundance depends only on the $r$-dependence of $T$. 
From Eqs~(\ref{Tvis}) and (\ref{Tirr}), the transition from $T_{\textrm{vis}}$
to $T_{\rm irr}$ occurs at
\begin{align}
    \label{eq:r_vis_irr}
    r & \simeq 7.1 
    \left(\frac{L_*}{1.0L_{\odot}}\right)^{-20/33}
    \left(\frac{M_*}{1.0M_{\odot}}\right)^{31/33}
    \left(\frac{\alpha}{10^{-2}}\right)^{-14/33}
    \left(\frac{\dot{M}_\textrm{g}}{10^{-7}M_{\odot}/\textrm{yr}}\right)^{28/33}
    \,\textrm{au}.
\end{align}

The gas disk aspect ratios corresponding to Eqs.~(\ref{Tvis}) and (\ref{Tirr}) are
\begin{align}
    \label{hvis}
    h_{\textrm{g,vis}} & =\frac{H_{\textrm{g,vis}}}{r}\simeq 0.035\left(\frac{M_*}{1.0M_{\odot}}\right)^{-7/20}\left(\frac{\alpha}{10^{-2}}\right)^{-1/10}\left(\frac{\dot{M}_\textrm{g}}{10^{-7}M_{\odot}/\textrm{yr}}\right)^{1/5}\left(\frac{r}{1\,\textrm{au}}\right)^{1/20}, \\
    \label{hirr} 
    h_{\textrm{g,irr}} & =\frac{H_{\textrm{g,irr}}}{r}\simeq 0.022\left(\frac{L_*}{1.0L_{\odot}}\right)^{1/7}\left(\frac{M_*}{1.0M_{\odot}}\right)^{-4/7}\left(\frac{r}{1\,\textrm{au}}\right)^{2/7}.
\end{align}
From the assumption of steady accretion, 
$\dot{M}_{\rm g} = 3\pi \Sigma_{\rm g} \alpha H_{\rm g}^2 \Omega$,
where $\Sigma_{\rm g}$ is the disk gas surface density and $\Omega$ is the Keplerian frequency,
the gas surface densities the viscous-heating 
and irradiation dominated regions are
\begin{align}
    \label{Svis}
    \Sigma_{\textrm{g,vis}} & \simeq 1.2\times 10^3\left(\frac{M_*}{1.0M_{\odot}}\right)^{1/5}\left(\frac{\alpha}{10^{-2}}\right)^{-4/5}\left(\frac{\dot{M}_\textrm{g}}{10^{-7}M_{\odot}/\textrm{yr}}\right)^{3/5}\left(\frac{r}{1\,\textrm{au}}\right)^{-3/5} \textrm{g cm}^{-2}, \\
    \label{Sirr}
    \Sigma_{\textrm{g,irr}} & \simeq 3.0\times 10^3\left(\frac{L_*}{1.0L_{\odot}}\right)^{-2/7}\left(\frac{M_*}{1.0M_{\odot}}\right)^{9/14}\left(\frac{\alpha}{10^{-2}}\right)^{-1}\left(\frac{\dot{M}_\textrm{g}}{10^{-7}\,M_{\odot}/\textrm{yr}}\right)\left(\frac{r}{1\,\textrm{au}}\right)^{-15/14} \textrm{g cm}^{-2}.
\end{align}
We use the gas disk aspect ratio \(h_{\textrm{g}}=H_{\textrm{g}}/r=\max(h_{\textrm{g,vis}},h_{\textrm{g,irr}})\) and the gas surface density \(\Sigma_{\textrm{g}}=\min(\Sigma_{\textrm{g,vis}},\Sigma_{\textrm{g,irr}})\).

We simply set the snow line by $T = 170\,{\rm K}$.
From Eqs.~(\ref{Tvis}) and (\ref{Tirr}), the snowline radius is given by \(r_{\textrm{snow}}=\max(r_{\textrm{snow,vis}},r_{\textrm{snow,irr}})\), where
\begin{align}
    r_{\textrm{snow,vis}} & \simeq 2.0\left(\frac{M_*}{1.0M_{\odot}}\right)^{1/3}\left(\frac{\alpha}{10^{-2}}\right)^{-2/9}\left(\frac{\dot{M}_\textrm{g}}{10^{-7}M_{\odot}/\textrm{yr}}\right)^{4/9}\left(\frac{T}{170\,\textrm{K}}\right)^{-10/9} \textrm{au},\\
    r_{\textrm{snow,irr}} & \simeq 0.53\left(\frac{L_*}{1.0L_{\odot}}\right)^{2/3}\left(\frac{M_*}{1.0M_{\odot}}\right)^{-1/3}\left(\frac{T}{170\,\textrm{K}}\right)^{-7/3} \textrm{au}.
\end{align}
The snow line is determined by viscous heating when \(\dot{M}_{\textrm{g}}\ga 5.0\times 10^{-9}\,M_{\odot}/\textrm{yr}\) for \(\alpha=10^{-2}\).
Similarly, we set the silicate line radius \(r_{\textrm{sil}}\) as the location of silicate sublimation temperature (\(T = 1400\,\)K), 
and the annealing line \(r_{\textrm{annl}}\) by $T = 1000\,{\rm K}$.
When the viscous heating dominates at the snow line, 
it also dominates at the silicate and annealing lines and
they are given by
\begin{align}
    r_{\rm annl} & \simeq 0.14\, r_{\rm snow} \\
    r_{\rm sil} & \simeq 0.096\, r_{\rm snow}. 
\end{align}
If the irradiation dominates, the radial dependence of $T$ is weaker and $r_{\rm annl} \simeq 0.016\, r_{\rm snow}$.

We use $M_*=M_{\odot}$ and $L_*=L_{\odot}$.
As a result, the disk parameters are only the disk accretion rate $\dot{M}_{\rm g}$ and
the viscosity parameter $\alpha$. 
The fiducial values of our disk model are $\dot{M}_\textrm{g}=10^{-7}M_{\odot}/\textrm{yr}$ and $\alpha=10^{-2}$.
We will also show the results with $\alpha=10^{-3}$.
Another important parameter is the pebble to gas mass flux ratio, 
$F_{\rm p/g} = \dot{M}_{\rm peb}/\dot{M}_{\rm g}$ (where $\dot{M}_{\rm peb}$ is the pebble mass flux).
Because pebbles start effective drift when they grow enough that the drift timescale becomes shorter than the growth timescale, the pebble mass flux is regulated by
the outward migration of the pebble formation front where the two timescales coincide \citep[][]{Lambrechts2014,Ida2016}.
\cite{Ida&Guillot2016} estimated that 
$F_{\rm p/g} \sim 0.03 (\alpha/10^{-2})^{-1} (t/10^6 {\rm yr})^{-8/21}$.
Because $F_{\rm p/g}$ depends on pebble growth model and disk structure,
we treat $F_{\rm p/g}$ as a parameter.
How to reflect the value of $F_{\rm p/g}$ to Monte Carlo simulation is
described in section \ref{subsec:sil_Sigma}.

\subsection{Monte Carlo simulation}
\label{subsec:MC}

\subsubsection{Overview}
\label{subsubsec:MC_overview}

\begin{figure}
    \plotone{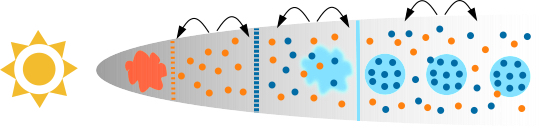}
    \caption{Overview of our models for gas disk and particle evolution in 
    different disk regions.
    The blue and orange particles represent amorphous and crystalline silicates. The light blue particles represent icy pebbles. 
The light blue solid, blue and orange dashed lines are the snow line, the annealing line and the silicate line. 
Icy pebbles enclosing many small amorphous silicates drift to the snow line (blue dashed line) from outer region with a certain accretion rate \(\dot{M}_{\textrm{peb}}\).
The amorphous/crystalline silicates are sticked to their surface with the given
probability (Eqs.~(\ref{prob}) and (\ref{eq:prob2})) beyond the snow line.
The icy mantle sublimates and the silicates in the icy pebbles are released. 
Inside the annealing line (the blue dashed line), amorphous precursors are annealed to become crystalline silicates. 
In a further inner region within the silicate line (the orange dashed line), 
crystalline silicates sublimate. 
There are only crystalline silicates inside annealing line and there are no particles inside silicate line. 
\label{fig:model}}
\end{figure}

Our models for gas disk and particle evolution in different disk regions
are illustrated in Figure \ref{fig:model}.
Pebbles are formed in outer disk regions and drift inward all through the disk due to gas drag.
When icy pebbles are formed, many small  ``amorphous" silicate dust particles (represented by 
dark blue dots) are embedded in the icy mantle (represented by light blue),
 because we assume that collisions between silicate and icy particles are as sticky as
ice-ice collisions (the threshold collision velocity for fragmentation, 
$v_{\rm frag}\ga 10\,{\rm m/s}$ \citep[e.g.,][]{Wada2013,Gundlach2015},
while silicate-silicate collisions are not sticky, 
$v_{\rm frag}\sim 1\,{\rm m/s}$ \citep[e.g.,][]{Wada2013}.
The Stokes number \(\tau_{\textrm{s,peb}}\) is given by Eq.~(\ref{eq:tau_peb}) ($\ga 0.1$). The typical size of pebbles is 10--100 cm (Eq.~(\ref{eq:R_peb})).

When the pebbles drift inward to $r<r_{\rm snow}$,
the ice mantle is sublimated and the embedded amorphous silicates are released.
During the drift, crystalline and amorphous silicate particles that diffused out from
inner regions (see below) also stick to the surface of drifting icy pebbles
with the probability derived in section \ref{subsec:stick}.
They are also re-released at the snow line.

The Stokes number of silicate particles is assigned to be \(\tau_{\textrm{s,sil}}\),
which we use $\tau_{\rm sil}=10^{-5}$, corresponding to the dust particle size of 
\(10 {\rm \mu m}\) for \(\alpha=10^{-2}\).
The radial drift velocity of the silicate particles is given by Eq.~(\ref{v_r}). 
It shows that the radial velocity of the silicate particles 
and the calculation results are independent of the detailed value of \(\tau_{\textrm{s,sil}}\)
(equivalently, the silicate particle size), 
as long as \(\tau_{\textrm{s,sil}} < \alpha\) (diffusion-dominated for the silicate particles) and \(\tau_{\textrm{s,peb}}> \alpha\) (drift-dominated for pebbles). 
We adopt a relatively severe rebounding/fragmentation limit
to neglect growth of silicate particles through mutual collisions. 
Even if the detailed silicate dust growth model is adopted, the
above condition would be satisfied for \(\alpha=10^{-3}\)--\(10^{-2}\) that we use in this paper. 
Detailed discussions on \(\tau_{\rm s,sil}\) is given in the Appendix. 

Once the silicates arrive at $r< r_{\rm annl}$,
they are identified as ``crystalline" silicates (represented by orange dots).
Even if they go back to the regions at $r > r_{\rm annl}$,
they keep the state of crystals, because crystalline silicates are stable.
When the silicates arrive at $r< r_{\rm sil}$, 
they should become gas molecules.
We assign $\tau_{\rm sil}=0$ for them in this region.
When they go back to the regions $r> r_{\rm sil}$, they re-condense
as crystalline silicates. 
About 5\% of the crystalline silicate particles diffuse out beyond the snow line
and eventually diffuse inward toward the star.
If the icy pebbles are still drifting there, the crystalline silicates attach to the pebbles 
to rapidly return to the snow line again.
Once the steady state is established, the silicate loss rate to the star
is balanced with their injection rate at the snow line, although individual
silicate particles have different radial (sometimes, rather complicated) travel paths. 
These silicate particles tracking is summarized in Table~\ref{tab:track}.

\subsubsection{Detailed Monte Carlo simulation methods}

\begin{table}
\begin{center}
\begin{tabular}{c|c|c}\hline
location & Stokes number & structure \\ \hline \hline 
$r > r_{\rm snow}$ &  $\tau_{\rm s,peb}$ (sticking) & amorphous/crystalline \\
                            &  $\tau_{\rm s,sil}$ (not sticking) & \\
\hline
$r_{\rm annl} < r < r_{\rm snow}$ &  $\tau_{\rm s,sil}$ & amorphous/crystalline \\
\hline
$r_{\rm sil} < r < r_{\rm annl}$ & $\tau_{\rm s,sil}$ & $\rightarrow$ crystalline \\
\hline
$r < r_{\rm sil}$ &  0  & $\rightarrow$ gas \\
\hline\hline
\end{tabular}
\end{center} 
\caption{Changes of silicates.  
If the particles stick to icy pebbles,
they drift/diffuse with the pebbles of \(\tau_{\textrm{s,peb}}\) given by Eq.~(\ref{eq:tau_peb}), which is usually $\ga 0.1$.
If not, they drift/diffuse with $\tau_{\rm s,sil} = 10^{-5}$.
At $r_{\rm sil} < r < r_{\rm annl}$, they change to crystalline silicate particles,
while they change to gas molecules at $r < r_{\rm sil}$.
At $r > r_{\rm annl}$, they do not change structure.}
\label{tab:track}
\end{table}

\begin{figure} 
\epsscale{0.7}
\plotone{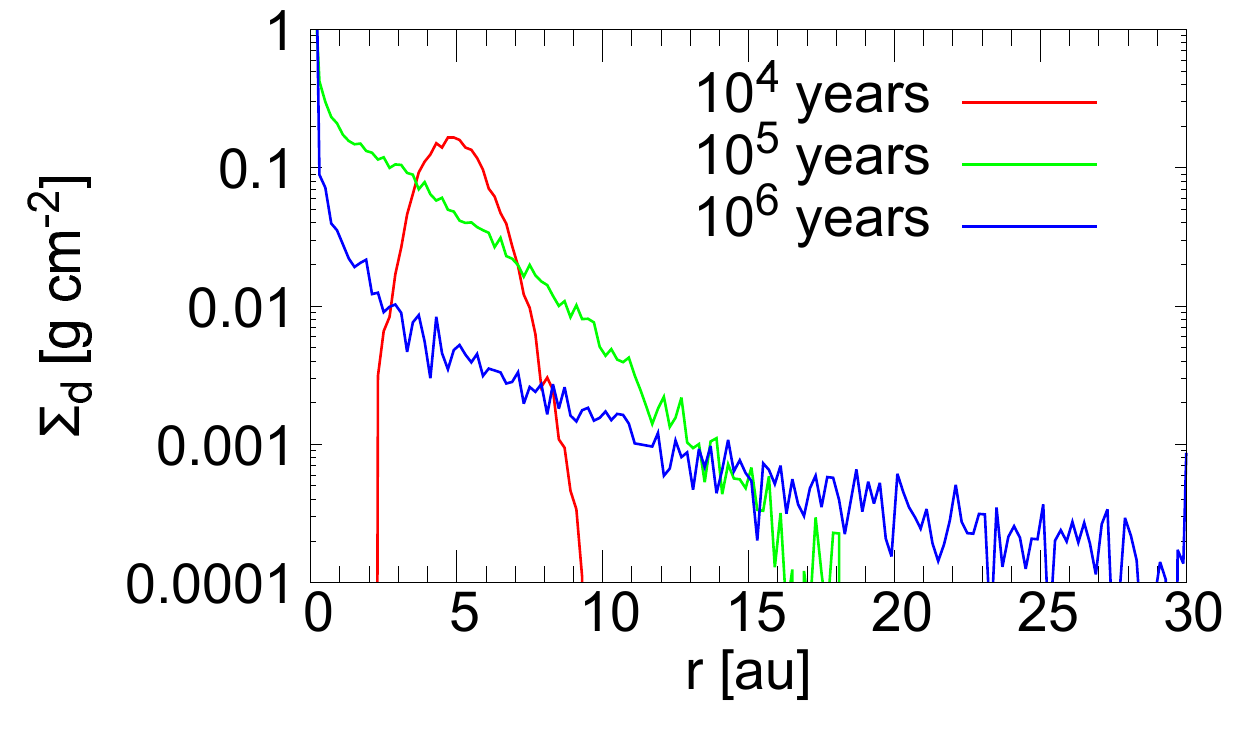} 
 \caption{Reproduction of Fig. 1 of  \cite{Ciesla2011} \label{fig:Ciesla}}
\end{figure}

In order to follow the orbital evolution of silicate particles described in
section \ref{subsubsec:MC_overview},  
we employ the 3D Monte Carlo simulation method for advection and diffusion 
of silicate particles developed by \cite{Ciesla2011}.
\cite{Ciesla2011} showed that the surface density evolution
of silicate particles determined by the concentration equation, 
\begin{equation}    \label{conti}    
\dfrac{\partial \Sigma_{\textrm{sil}}}{\partial t}=\dfrac{1}{r}\dfrac{\partial }{\partial r}\left[r\Sigma_{\textrm{sil}}\left(v_{r}-\frac{D}{\Sigma_{\textrm{sil}}/\Sigma_{\textrm{g}}}\frac{\partial}{\partial r}\left(\frac{\Sigma_{\textrm{sil}}}{\Sigma_{\textrm{g}}}\right)\right)  \right],
\end{equation} 
where $\Sigma_{\textrm{sil}}$ is the surface density of the silicate particles, $D$ is their diffusivity, and $v_r$ is their radial advection velocity, is reproduced by the following 3D Monte Carlo simulations of many particles.
The changes in the positions in Cartesian coordinates ($x, y, z$) of each particle after the timestep $\delta t$ are given by
\begin{align}
    \delta x & =v_x \delta t +\mathcal{R}_x\sqrt{6D\delta t}, \label{eq:diffx}\\
    \delta y & =v_y \delta t +\mathcal{R}_y\sqrt{6D\delta t}, \label{eq:diffy}\\
    \delta z & =v_z \delta t +\mathcal{R}_z\sqrt{6D\delta t}, \label{eq:diffz}
\end{align}
where 
$\delta t$ is given by the inverse of the local Keplerian frequency,
$\delta t = \Omega^{-1}$, and 
the first and second terms in the right hand sides represent
the advection and diffusion.
In the diffusion terms, \(\mathcal{R}_x\), \(\mathcal{R}_y\) and \(\mathcal{R}_z\) are independent random numbers in a range of [-1,1]
with a root mean square of $1/\sqrt{3}$.
Because $\tau_{\rm s,sil} \ll 1$ and the back-reaction due to silicate particle concentration 
is neglected (assuming the silicate to gas density $\rho_{\rm sil} / \rho_{\rm g} \ll 1$) in this paper, we identify the diffusivity of the particles, $D$,
as the gas turbulent viscosity \citep[][]{Hyodo2019}, 
\begin{align}
\nu = \alpha h_{\rm g}^2 r^2 \Omega. 
\end{align}
The advection velocities are given by \citep{Ciesla2010,Ciesla2011},
\begin{align}
    \label{v_x}
    v_{x} & =\left[v_{\textrm{drag},r}+\frac{1}{\Sigma_{\textrm{g}}}\frac{\partial(D\Sigma_{\textrm{g}})}{\partial r}\right]\times\frac{x}{r}, \\
    \label{v_y}
    v_{y} & =\left[v_{\textrm{drag},r}+\frac{1}{\Sigma_{\textrm{g}}}\frac{\partial(D\Sigma_{\textrm{g}})}{\partial r}\right]\times\frac{y}{r}, \\
    \label{v_z}
    v_{z} & =v_{\textrm{drag},z}+\frac{1}{\rho_{\textrm{g}}}\frac{\partial(D\rho_{\textrm{g}})}{\partial z}.
\end{align}
In the steady accretion gas disk
($\partial \dot{M}_{\rm g}/\partial r
= \partial (3\pi \Sigma_{\rm g} \nu)/\partial r = 0$),
the second terms in Eqs.~(\ref{v_x}) and (\ref{v_y}) vanish. 
Assuming the vertically isothermal disk with gas density \(\rho_{\textrm{g}} \propto \exp(-z^2/2H_{\textrm{g}}^2)\), 
the second term in Eq.(\ref{v_z}) is \(-\alpha\Omega z\).
The radial drift velocity due to gas drag is given by
\citep[][]{Ida&Guillot2016,2017Schoonenberg}
\begin{equation}   \label{v_r0}    
v_{{\rm drag},r} \simeq -\frac{1}{1+\tau^2_{\rm s,sil}}
\left(2\tau_{\rm s,sil} \eta v_{\textrm{K}}+u_{\nu}\right),
\end{equation}
where $v_{\rm K}$ is the local Keplerian velocity, \(u_{\nu}\) is the disk gas (inward) accretion velocity given by $u_{\nu} = 3\nu/2r=(3\alpha h^2_{\textrm{g}}/2)v_{\textrm{K}}$,
and \(\eta\) is the degree of deviation of the gas rotation angular velocity \(\Omega\) from Keplerian one, given by
\begin{equation}
    \eta\equiv\frac{\Omega_{\textrm{K}}-\Omega}{\Omega_{\textrm{K}}}=
    \frac{1}{2}\left|\frac{d\ln P}{d\ln r}\right| h_{\textrm{g}}^2 =
    C_{\eta}h_{\textrm{g}}^2.
    \label{eq:mu}
\end{equation}
Hereafter we approximate $1/(1+\tau_{\rm s,sil}^2)$ as $\sim 1$, because $\tau_{\rm s,sil} \ll 1$.
Substituting the expression of $u_\nu$ and Eq.~(\ref{eq:mu}) into Eq.~(\ref{v_r0}),
we obtain
\begin{align}
\label{v_r}  
v_{\textrm{drag},r}= - \left(2 C_{\eta} \tau_{\textrm{s,sil}}+\frac{3}{2}\alpha\right)h^2_{\rm g}v_{\rm K}.
\end{align}
For simplicity, we take \(C_{\eta}=11/8\) for $T\propto r^{-1/2}$. 
The vertical drift velocity \(v_{\textrm{drag},z}\) is given by \citep{Ciesla2010}
\begin{equation}
    v_{\textrm{drag},z}=-\tau_{\textrm{s,sil}}\Omega z. \label{eq:v_dragz}
\end{equation}
Substituting Eqs.~(\ref{v_x}) to (\ref{eq:v_dragz}) into
Eqs.~(\ref{eq:diffx}) to (\ref{eq:diffz}), we obtain
\begin{align}
    \delta x & =- \left(2 C_{\eta} \tau_{\textrm{s,sil}}+\frac{3}{2}\alpha\right)h^2_\textrm{g}x +\mathcal{R}_x\sqrt{6\alpha}h_{\textrm{g}}r, \\
    \delta y & =- \left(2 C_{\eta} \tau_{\textrm{s,sil}}+\frac{3}{2}\alpha\right)h^2_\textrm{g}y +\mathcal{R}_y\sqrt{6\alpha}h_{\textrm{g}}r, \\
    \delta z & =-(\tau_{\textrm{s,sil}}+\alpha)z+\mathcal{R}_z\sqrt{6\alpha}h_{\textrm{g}}r.
\end{align}
We confirmed that
these equations reproduce the evolution of a Gaussian-ring 
\citep[][Fig. 1]{Ciesla2011}, as shown in Fig.~\ref{fig:Ciesla} in our paper.

We inject super-particles that represent a swarm of silicate particles
and suffer specific drag force for realistic particles near the snow line at every timestep 
\(\delta t_{\textrm{inj}}\). 
The initial radii are randomly distributed
in the range of \([r_{\textrm{snow}}-0.5\Delta r_0, r_{\textrm{snow}}+0.5\Delta r_0]\) and the vertical height is determined by a Gaussian distribution of the root mean square \(\Delta z_0\). 
Following \cite{Ida2021}, we adopt \(\Delta r_0=0.1\, H_{\textrm{g}}\) and \(\Delta z_0=H_{\textrm{p}}\) where \(H_{\textrm{p}}\) is a scale-hight of pebbles given by 
\citep[][]{Dubrulle1995,Youdin2007}
\begin{equation}
    H_{\textrm{p}}=\sqrt{\frac{\alpha}{\alpha+\tau_{\textrm{s,peb}}}}H_{\textrm{g}}.
\end{equation} 
Shorter (longer) \(\delta t_{\rm inj}\) corresponds to a larger (smaller) number of super-particles
with relatively smaller (larger) individual masses in the simulations. 
We choose \(\delta t_{\rm inj}\) short enough to statistically calculate the silicate surface densities and large enough for numerical simulations.

When silicate particles enter the region at $r < 0.01r_{\rm snow}$,
we regard that the particles are lost to the central star and remove them
from the simulations.

\subsection{Surface density of particles}
\label{subsec:sil_Sigma}

If we are concerned only with relative abundance between
amorphous and crystalline silicates at each $r$, we do not need the information 
of the surface density of silicate particles.
However, to calculate the collision probability of silicate particles with pebbles
at $r > r_{\rm snow}$, we need to specify the surface density of the pebbles.
Accordingly, we can also calculate the silicate particle surface density, $\Sigma_{\rm sil}$.
The surface density of the silicate particles 
is calculated by the mass of individual silicate (super-)particles
given from the pebble mass flux 
by the method of \cite{Ida2021}, as shown below.

The silicate injection mass rate is given by
\begin{equation}
    \dot{M}_{\rm sil} = f_{\rm sil} \, \dot{M}_{\textrm{peb}} = f_{\rm sil} \, F_{\textrm{p/g}}\dot{M}_{\textrm{g}},
\end{equation}
where \(f_{\textrm{sil}}\) is the silicate mass fraction in drifting pebbles (\(f_{\textrm{sil}}=0.5\) in the normal case) and \(\dot{M}_{\textrm{g}}\) is the disk gas accretion rate. 
The mass of an individual super-particle, \(m\), is given by
\begin{equation}
  \dot{M}_{\textrm{sil}}=m/\delta t_{\textrm{inj}}.
\end{equation}
The surface density of silicate particles at $r$ is given 
with the number of particles \(\Delta N_r\) in the radial width 
of $\Delta r$ around $r$ by
\begin{equation}
    \Sigma_{\textrm{sil}}= \frac{m\Delta N_r}{2\pi r\Delta r}=\frac{\Delta N_r \dot{M}_{\textrm{sil}}\delta t_{\textrm{inj}}}{2\pi r^2\times 2.3(\Delta\log_{10}r)}
= 0.65 \, h_{\rm g}^2 \,\alpha (\delta t_{\textrm{inj}}\Omega)
f_{\textrm{sil}} F_{\rm p/g} \frac{\Delta N_r}{\Delta \log_{10}r} \times \Sigma_{\rm g},
\label{eq:sigcal}
\end{equation}
where we used $\Sigma_{\textrm{g}}=\dot{M}_{\textrm{g}}/2\pi ru_{\nu}
=\dot{M}_{\textrm{g}}/3\pi\alpha H^2_{\textrm{g}}\Omega$.


\subsection{Sticking probability to icy pebbles}
\label{subsec:stick}

While we neglect silicate particle growth due to silicate-silicate collisions, we include the effect of 
coagulation of silicate particles with pebbles, assuming that silicate-ice collisions are more sticky. By 
the sticking of diffused-out silicate particles to drifting pebbles at \(r>r_{\rm snow}\), the silicate components are 
quickly returned to the disk inner region \citep{Misener2019}. 
We will show later that this effect is very important 
for the distributions of silicates beyond the snow line.

Assuming the perfect sticking of silicate particles to icy pebbles, 
the probability \(P\) for a silicate particle to collide with any of pebbles
during $\Delta t$, which we call ``sticking probability," is calculated by  
\begin{equation}
    P=n_{\rm peb} \sigma v\Delta t = 
   \frac{H_{\rm peb}}{H_{\rm g}} \, n_{\rm peb}(\pi R_{\rm peb}^2)v\Delta t,
    \label{eq:sticking0}
\end{equation}
where \(n_{\rm peb}\) is the number density of the pebbles, \(R_{\rm peb}\) is the physical radius of the pebbles, and \(v\) is the relative velocity between a silicate dust particle and a pebble.
For $\alpha \ga 10^{-3}$, it is dominated by turbulent mixing and is given by 
\citep{Sato2016}
\begin{align}
v \simeq \sqrt{ 3\alpha (\tau^2_{\textrm{s,sil}}+\tau^2_{\textrm{s,peb}})^{1/2}} \, c_s
  \simeq\sqrt{3\alpha \, \tau_{\rm s,peb}} \, H_{\textrm{g}}\Omega
\label{eq:vd_turb}.
\end{align}
The factor, $H_{\rm peb}/H_{\rm g}$, in Eq.~(\ref{eq:sticking0}) reflects that
the scale height of the pebbles, $H_{\rm peb}$, is smaller than
that of silicate dust particles, $H_{\rm d} \sim H_{\rm g}$,
and the collisions are possible only during the silicate particles
are passing the region of $\mid z \mid < H_{\rm peb}$.
The number density of pebbles is  
\begin{equation}
    n_{\rm peb} 
 =\frac{\Sigma_{\textrm{peb}}}{\sqrt{2\pi}H_{\textrm{peb}}}
 \frac{1}{(4\pi/3)\rho_{\textrm{bulk}} R_{\rm peb}^3},
\end{equation}
where $\Sigma_{\rm peb}$ is the surface density of pebbles.
Because the scale height of the pebbles in the above equation, $H_{\rm peb}$, 
cancels out with the factor in Equation~(\ref{eq:sticking0}), we obtain
\begin{equation}
    P=  
    \frac{3}{4\sqrt{2\pi}} \frac{\Sigma_{\textrm{peb}}}{H_{\textrm{g}}}
 \frac{v\Delta t}{\rho_{\textrm{bulk}} R_{\rm peb}}
 = \frac{3}{4} \sqrt{\frac{3}{2\pi}}
 \frac{\Sigma_{\textrm{peb}}}{\rho_{\textrm{bulk}} R_{\rm peb}}
  \, \sqrt{\alpha \, \tau_{\rm s, peb}}\,\Omega \Delta t,
    \label{eq:sticking}
\end{equation}
where we used Eq.~(\ref{eq:vd_turb}).

We assume that the bulk density of icy pebbles is \(\rho_{\textrm{bulk}}\simeq 1.0\,\textrm{g cm}^{-3}\). 
From Eq.(\ref{v_r}) with replacing $\tau_{\rm s,sil}$ by $\tau_{\rm s,peb}$
and \(\alpha\ll\tau_{\textrm{s,peb}}\), 
\begin{equation}
    \Sigma_{\textrm{peb}}
\simeq \frac{\dot{M}_{\textrm{peb}}}{2\pi r\times 2C_{\eta}\tau_{\textrm{s,peb}}h_{\rm g}^2 v_{\rm K}} =\frac{\dot{M}_{\textrm{peb}}}{4\pi C_{\eta}\tau_{\textrm{s,peb}}H^2_{\textrm{g}}\Omega}
\simeq\frac{3\alpha F_{\rm p/g}}{4C_{\eta}\tau_\textrm{s,peb}} \Sigma_{\rm g},
\label{eq:Sigpeb1}
\end{equation}
where we used $\dot{M}_{\rm g} = 3 \pi \Sigma_{\rm g} \alpha H_{\rm g}^2 \Omega$.
In the Stoke regime
\footnote{In outer regions far beyond the snow line,
the Epstein drag is applied and the pebble size is smaller.
However, as we will show in section~\ref{sec:resu},
silicate particles cannot diffuse out so much to the outer regions
when the collisions with pebbles are efficient.
They can significantly spread only after pebbles are significantly depleted.
In that case, while the Epstein drag must be applied,
the collision effect is not important.
Thereby we only use the expression in the Stokes regime. 
}, 
the radial drift dominates over collisional growth for pebbles, and the pebble size is approximated by an \(r\)-independent form \citep{Ida2016},
\begin{equation}
    \label{eq:R_peb}
    R_{\rm peb} \simeq 60\left(\frac{\alpha}{10^{-2}}\right)^{1/20}\left(\frac{\dot{M}_\textrm{g}}{10^{-7}M_{\odot}/\textrm{yr}}\right)^{0.27}\left(\frac{F_{\textrm{p/g}}}{0.1}\right)^{0.32}\,\textrm{cm}.
\end{equation}
The corresponding Stoke number of pebbles is given by \citep{Ida2016}
\begin{equation}
    \label{eq:tau_peb}
    \begin{split}
        \tau_{\textrm{s,peb}}&\simeq 4.4\times 10^{-4}\left(\frac{\alpha}{10^{-2}}\right)^{1/10}\left(\frac{\dot{M}_\textrm{g}}{10^{-7}M_{\odot}/\textrm{yr}}\right)^{-1/5}\left(\frac{r}{1\,\textrm{au}}\right)^{-21/20}\left(\frac{R_{\rm peb}}{1\,\textrm{cm}}\right)^2\\
        &\simeq 1.6 \left(\frac{\alpha}{10^{-2}}\right)^{1/5}\left(\frac{\dot{M}_\textrm{g}}{10^{-7}M_{\odot}/\textrm{yr}}\right)^{0.34}\left(\frac{r}{1\,\textrm{au}}\right)^{-21/20}\left(\frac{F_{\textrm{p/g}}}{0.1}\right)^{0.64}.
    \end{split}
\end{equation}
Substituting Eqs.~(\ref{Svis}) and (\ref{eq:tau_peb}) into Eq.~(\ref{eq:Sigpeb1}),
\begin{equation}
    \Sigma_{\textrm{peb}}\simeq 0.41\left(\frac{\dot{M}_\textrm{g}}{10^{-7}M_{\odot}/\textrm{yr}}\right)^{0.26}\left(\frac{r}{1\,\textrm{au}}\right)^{9/20}\left(\frac{F_{\textrm{p/g}}}{0.1}\right)^{0.36}\,\textrm{g cm}^{-2}.
\end{equation}
Therefore, the sticking probability (Eq.~(\ref{eq:sticking})) is given by
\begin{equation}
    \label{prob}
        P 
\simeq 0.52 \, \frac{\Sigma_{\textrm{peb}}}{\rho_{\textrm{bulk}}R_{\rm peb}}\, \sqrt{\alpha \, \tau_{\textrm{s,peb}}}\, \Omega\Delta t
\simeq\zeta_{P} \, \Omega\Delta t,
\end{equation}
where
\begin{equation}
\label{eq:prob2}
    \zeta_{P}\simeq 4.5\times 10^{-4}\left(\frac{\alpha}{10^{-2}}\right)^{11/20}\left(\frac{\dot{M}_\textrm{g}}{10^{-7}M_{\odot}/\textrm{yr}}\right)^{0.16}\left(\frac{r}{1\,\textrm{au}}\right)^{-0.075}\left(\frac{F_{\textrm{p/g}}}{0.1}\right)^{0.36}.
\end{equation}

\section{Analytical solution of silicate surface density} \label{sec:analy}

In order to understand the results of our Monte Carlo simulations
presented in section \ref{sec:resu}, we here derive analytical steady solutions to 
the advection-diffusion equation of concentration given by Eq.(\ref{conti}). 
We also derive modifications due to the effect of the sticking to pebbles.

\subsection{Simple case}
\label{subsec:simple}

Equation (\ref{conti}) has two possible solutions in a steady state (\(\partial \Sigma_{\textrm{sil}}/\partial t=0\)),
\begin{align}
    \label{eq:ana1}
    \frac{\partial \dot{M}_{\textrm{sil}}}{\partial r} & =0, \\
    \label{eq:ana2}
    \dot{M}_{\textrm{sil}} & =0,
\end{align} 
where
\begin{equation}
    \dot{M}_{\textrm{sil}}=2\pi r\Sigma_{\textrm{sil}}v_{r,\textrm{sil}}\left(1-\frac{D/r}{v_{r,\textrm{sil}}}\frac{\partial\ln(\Sigma_{\textrm{sil}}/\Sigma_{\textrm{g}})}{\partial\ln r}\right).
\end{equation}
The former solution represents the inwardly accreting steady flow,
\begin{align}
& \frac{\partial \ln(\Sigma_{\rm sil}/\Sigma_{\rm g})}{\partial\ln r} = 0,  
\label{eq:Mdot_sil_Sigma}\\
& \dot{M}_{\textrm{sil}} = 2\pi r\Sigma_{\textrm{sil}}v_{r,\textrm{sil}}.
\label {eq:Mdot_sil}
\end{align}
Equation~(\ref{eq:Mdot_sil_Sigma}) means $\Sigma_{\rm sil}(r) \propto \Sigma_{\rm g}(r)$.
Assuming $D=\nu$ and $v_{r,\rm{sil}} \simeq - (3/2)(\nu/r)$ with \(\tau_{\rm s,sil} \ll\alpha\), 
 Eq.~(\ref{eq:Mdot_sil}) with $\dot{M}_{\textrm{sil}} = f_{\rm sil} F_{\rm p/g} \dot{M}_{\rm g}$ shows
\begin{equation}
\Sigma_{\rm sil}(r) = f_{\rm sil} F_{\rm p/g} \Sigma_{\rm g}(r).
\label{eq:Sig_sil_steady}
\end{equation}
The latter (Eq.~\ref{eq:ana2}) is the solution of zero net flux, given by
\begin{equation}
    \label{eq:Sigsil0}
    \frac{\partial\ln(\Sigma_{\textrm{sil}}/\Sigma_{\textrm{g}})}{\partial\ln r}=\frac{rv_{r,{\textrm{sil}}}}{D}=-\frac{3}{2},
\end{equation}
which means that the radial gradient is steeper for $\Sigma_{\rm sil}$ than for $\Sigma_{\rm g}$. 
This dependence of \(r\) is equivalent to Eq.(23) with $\rm Sc = 1$ 
($D = \nu$) in \cite{Pavlyuchenkov2007}.

Setting the radial gradient of gas and dust surface densities \(\Sigma_{\textrm{g}}\propto r^{-p}\) and \(\Sigma_{\textrm{sil}}\propto r^{-q}\), 
the former and latter solutions are 
\begin{align}
q = & \ p \hspace{15mm}  [\textrm{for inward steady accretion solution}], \label{eq:steady_index} \\
q = & \ p+3/2 \hspace{6mm} [\textrm{for zero net flux solution}].\label{eq:zero_net_index} 
\end{align}
In the viscous-heating and irradiation dominated regions,
$p = 3/5$ (Eq.~\ref{Svis}) and $p = 15/14$ (Eq.~\ref{Sirr}), respectively. 
We will show in section~\ref{subsec:nopeb} that
the silicate distributions inside and outside the snow line
correspond to $q=p$ and \(q=p+3/2\), respectively.

\subsection{Case with sticking to icy pebbles}

Once silicate particles stick to icy pebbles,
they inwardly drift with relatively high velocity as a part of the pebbles. 
If the sticking is considered, 
Eq.~(\ref{conti}) should additionally have an extra term at $r>r_{\rm snow}$.
Because it is no more exactly solved, we approximately 
estimate the modification to $\Sigma_{\rm sil}$ due to the sticking. 

The sticking to icy pebbles is a sink of silicates beyond snow line.
The sticking rate is given by \(\zeta_P\Omega\) in Eq.~(\ref{prob}). Thus, 
the decay rate of silicate outward diffusion flux 
due to the sticking is 
\begin{equation}
    \frac{d\dot{M}_{\textrm{sil,out}}}{d\Delta t}\simeq-\zeta_{P}\Omega\dot{M}_{\textrm{sil,out}},
    \label{eq:mdot_sil_out}
\end{equation}
where $\Delta t = t - t_0$ and $t_0$
is the time at which the diffusion flux departs from the snow line. 
Typical diffusion length from $r_{\rm snow}$ at $t$ is
\begin{equation}
    \label{deltar}
    \Delta r=\sqrt{<\mathcal{R}_r^2>}\sqrt{6\alpha H_{\textrm{g}}^2\Omega\Delta t}=\sqrt{2\alpha H_g^2\Omega\Delta t}.
\end{equation}
The solution to Eq.~(\ref{eq:mdot_sil_out}) corresponds to
\begin{equation}
    \label{Msil0}
\ln \left(\frac{\dot{M}_{\textrm{sil,out}}}{\dot{M}_{\textrm{sil,out,0}}}\right)
\simeq -\zeta_{P}\Omega\Delta t
\simeq -\frac{\zeta_{P}}{2\alpha h_g^2}\left(\frac{\Delta r}{r_{\rm snow}}\right)^2
\equiv -\gamma,
\end{equation}
where \(\dot{M}_{\textrm{sil,out,0}}\) is the silicate outward flux at the snow line.
Solving this equation in terms of $\Delta r/r_{\rm snow}$,
\begin{equation}
    \frac{\Delta r}{r}\simeq \sqrt{\frac{2\gamma\alpha h_g^2}{\zeta_P}}\simeq 0.23\gamma^{1/2}\zeta_r,
\end{equation}
where
\begin{equation}
    \label{eq:zetar}
    \zeta_r=\left(\frac{\alpha}{10^{-2}}\right)^{0.125}\left(\frac{\dot{M}_\textrm{g}}{10^{-7}M_{\odot}/\textrm{yr}}\right)^{0.12}\left(\frac{r}{1\,\textrm{au}}\right)^{0.0875}\left(\frac{F_{\textrm{p/g}}}{0.1}\right)^{-0.18},
\end{equation}
we substituted $\zeta_P$ given by Eq.~(\ref{eq:prob2}), 
and \(h_{\textrm{g}}\) is given by Eq.(\ref{hvis}) since viscous heating dominates at the snow line when \(\dot{M}_{\textrm{g}} \ga 1.5\times 10^{-9}\, M_{\odot}/{\textrm{yr}}\) for \(\alpha=10^{-3}\).

The effect of the decay in $\dot{M}_{\textrm{sil,out}}$ can be
expressed as modulation to $q = p+3/2$ for the zero net flux solution
in the simple case in section \ref{subsec:simple}.
The additional power-law index would be 
\begin{equation}
    q' = - \frac{\partial \ln (\dot{M}_{\textrm{sil,out}}/\dot{M}_{\textrm{sil,out,0}})}{\partial\ln r}
    \simeq - \frac{\ln (\dot{M}_{\textrm{sil,out}}/\dot{M}_{\textrm{sil,out,0}})}{\ln((r_{\textrm{snow}}+\Delta r)/r_{\textrm{snow}})}
    = \frac{\gamma}{\ln(1+0.23\gamma^{1/2}\zeta_r)}.
\end{equation}
The modified power-law index is given by
\begin{equation}
    \label{eq:Sigwipeb}
    q = p+\frac{3}{2}+ q'
       \simeq p+\frac{3}{2} + \frac{1}{\ln(1+0.23\,\zeta_r)},
\end{equation}
where we adopt $\gamma \simeq 1$. 

For $\dot{M}_{\rm g}=10^{-7}M_\odot/{\rm y}$, $\alpha = 10^{-2}$,
and $F_{\rm p/g}=0.1$, we obtain
$\zeta_r = 1.1$ and $q' = 4.4$ at 3 au.
In the viscous heating dominated regime, $q = 6.5$. 
For $\alpha = 10^{-3}$,
we obtain $\zeta_r = 0.83$, $q' = 5.7$, and $q=7.8$ at 3 au.
In the irradiation dominated regime ($p=15/14$),
$q$ is higher by 0.47.
Because $\zeta_r$ is insensitive to the disk parameters
(Eq.~\ref{eq:zetar}), the modified values of $q$ are not 
sensitively affected by the disk parameters. 
In section 4.2, we will show 
that Eq.~(\ref{eq:Sigwipeb}) reproduces the results of
the Monte Carlo simulations.

\section{Results} \label{sec:resu}

\subsection{Simple case} 
\label{subsec:nopeb}

\begin{figure*}
    \gridline{\fig{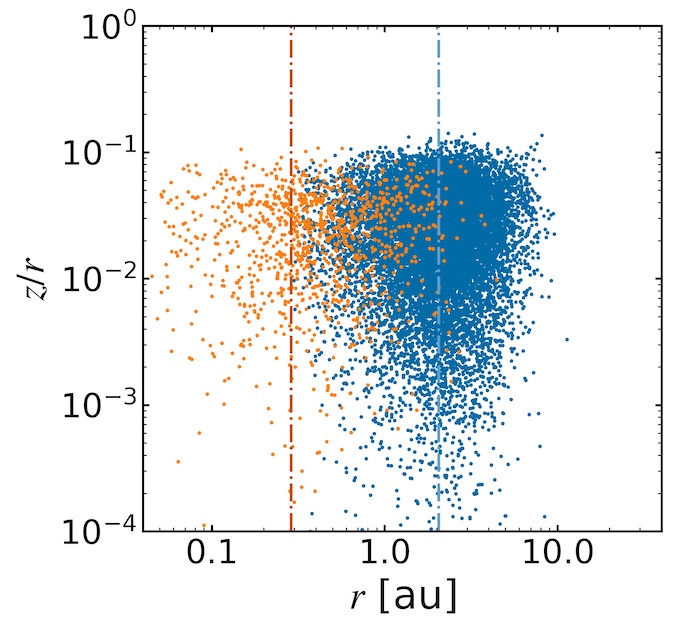}{0.25\textwidth}{(a)}
              \fig{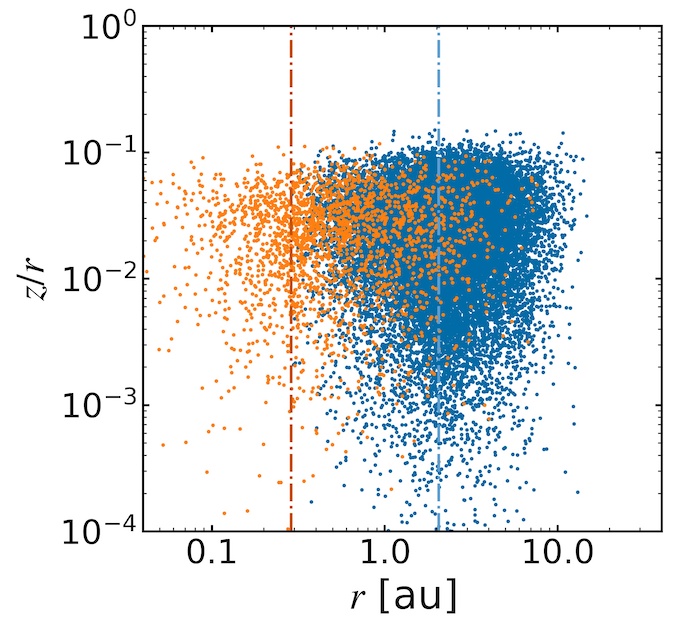}{0.25\textwidth}{(b)}
              \fig{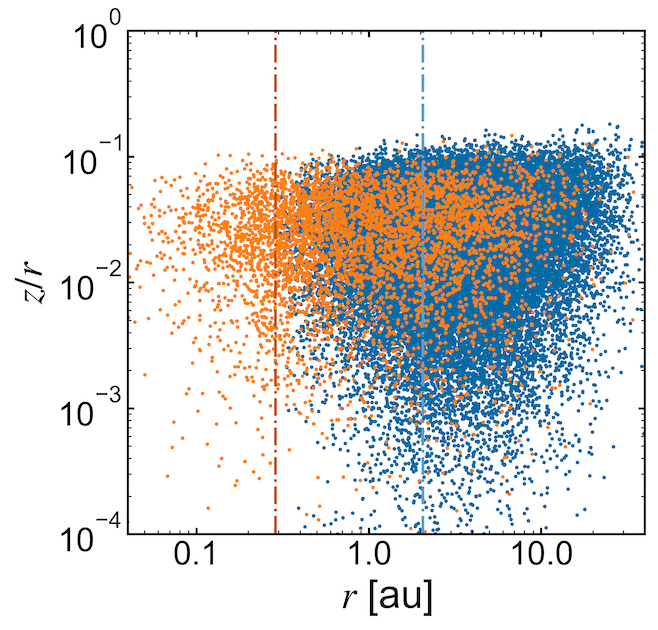}{0.25\textwidth}{(c)}
              \fig{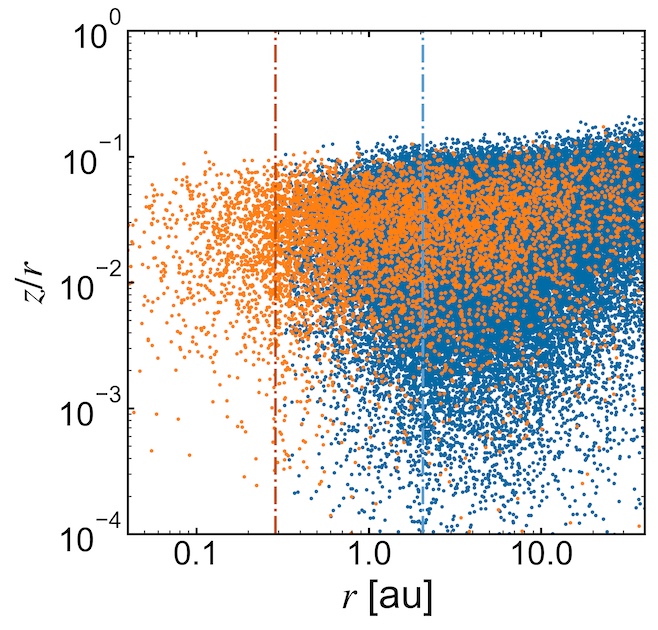}{0.25\textwidth}{(d)}
              }
    \gridline{\fig{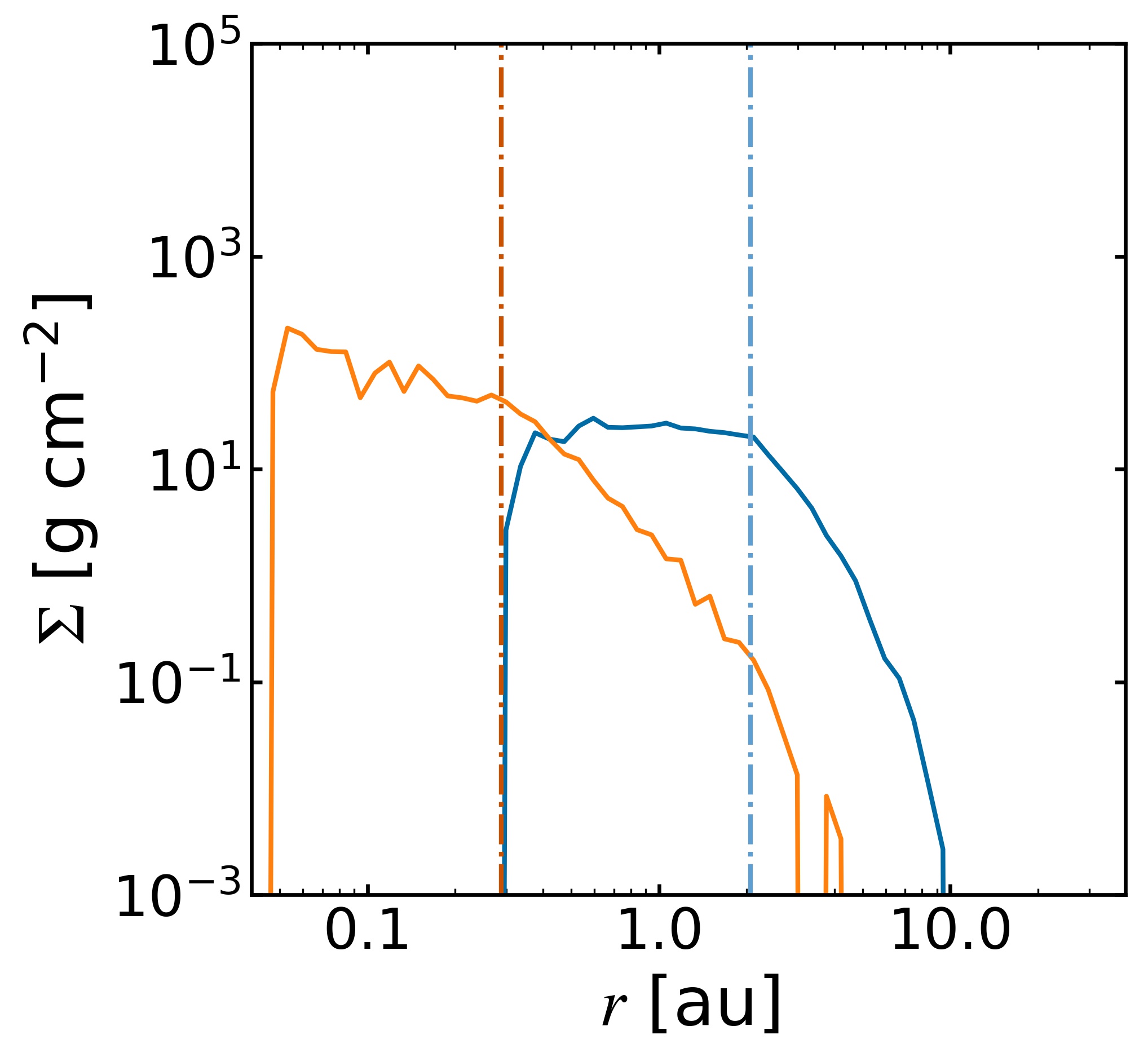}{0.25\textwidth}{(e)}
              \fig{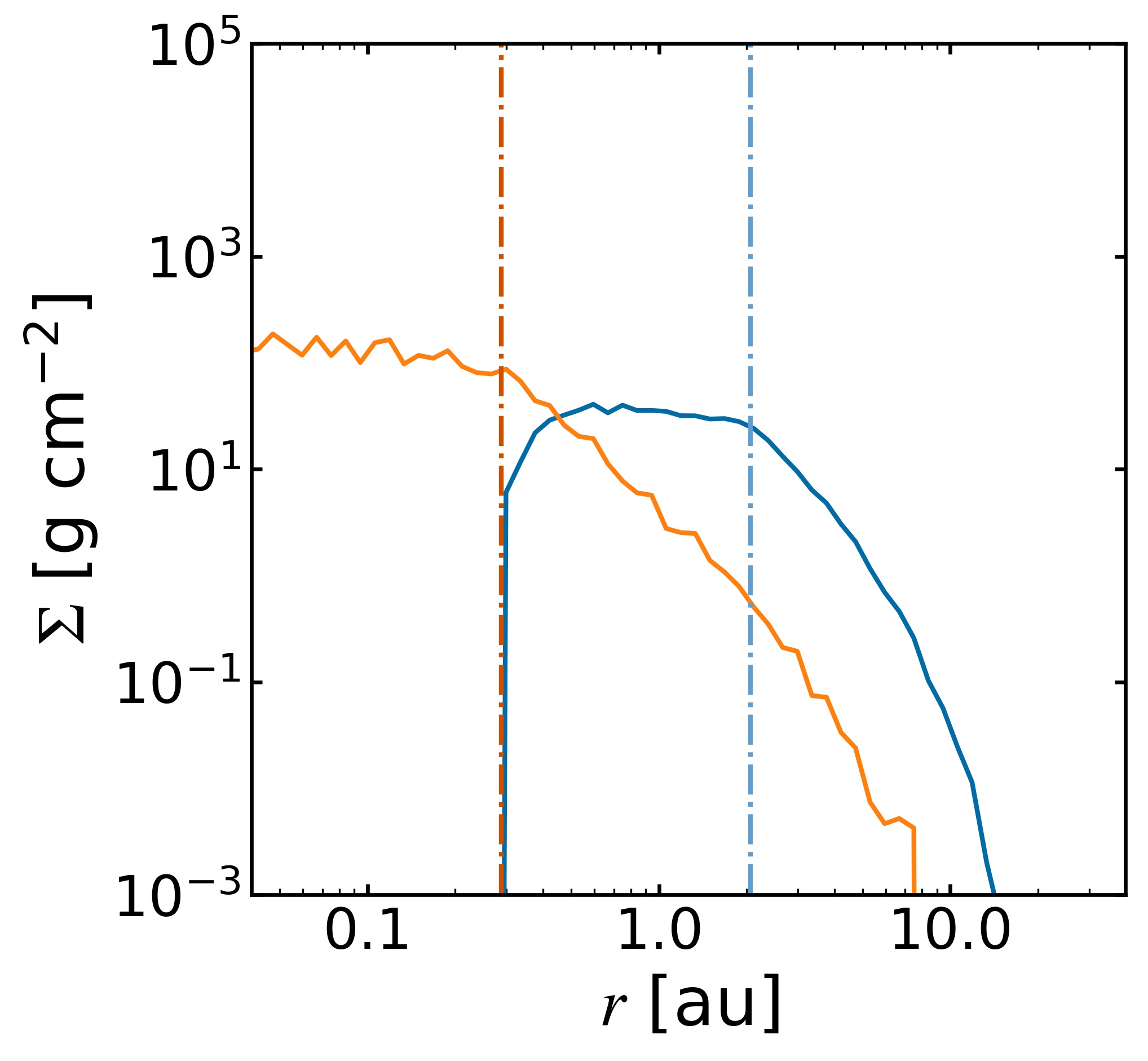}{0.25\textwidth}{(f)}
              \fig{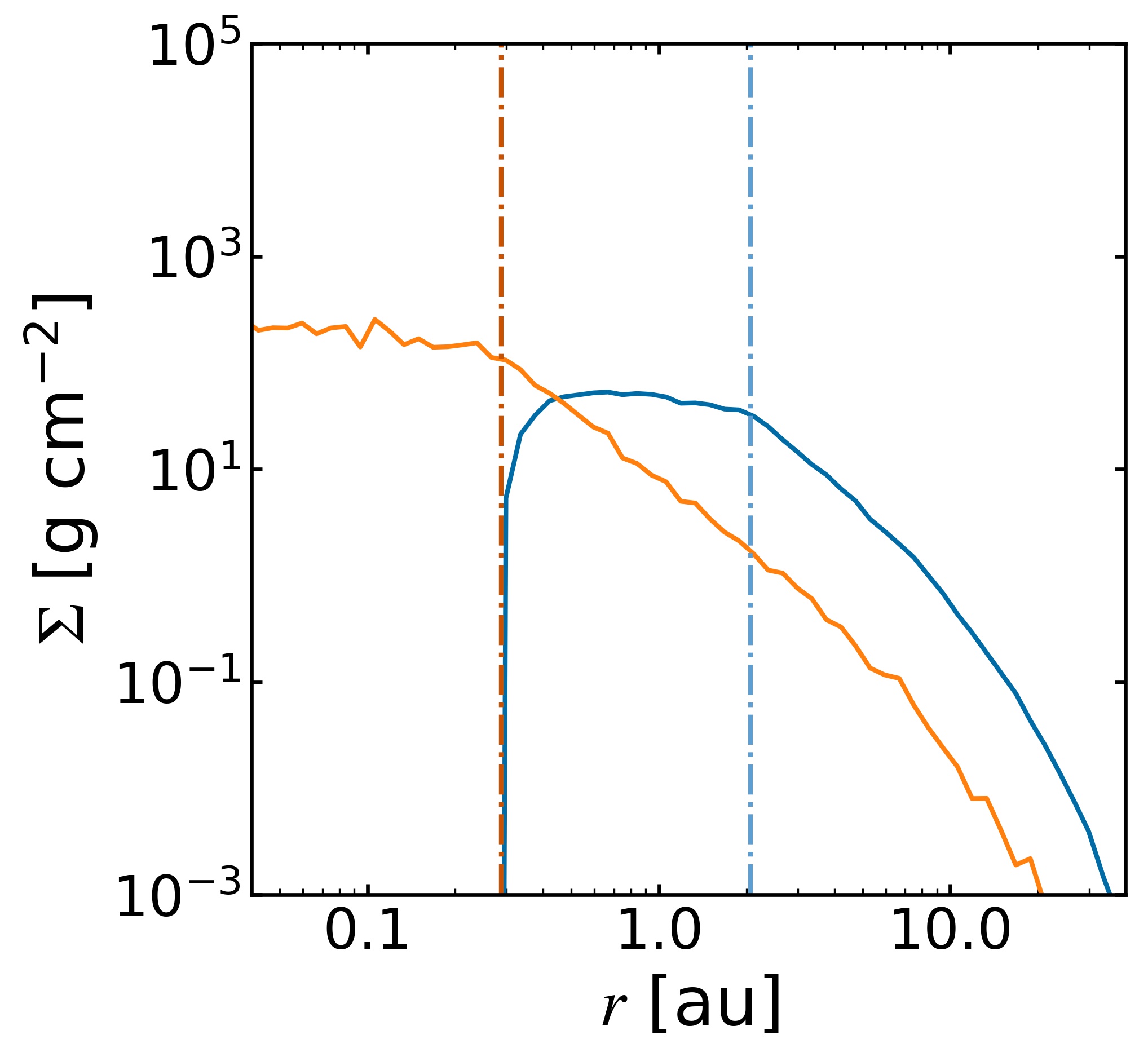}{0.25\textwidth}{(g)}
              \fig{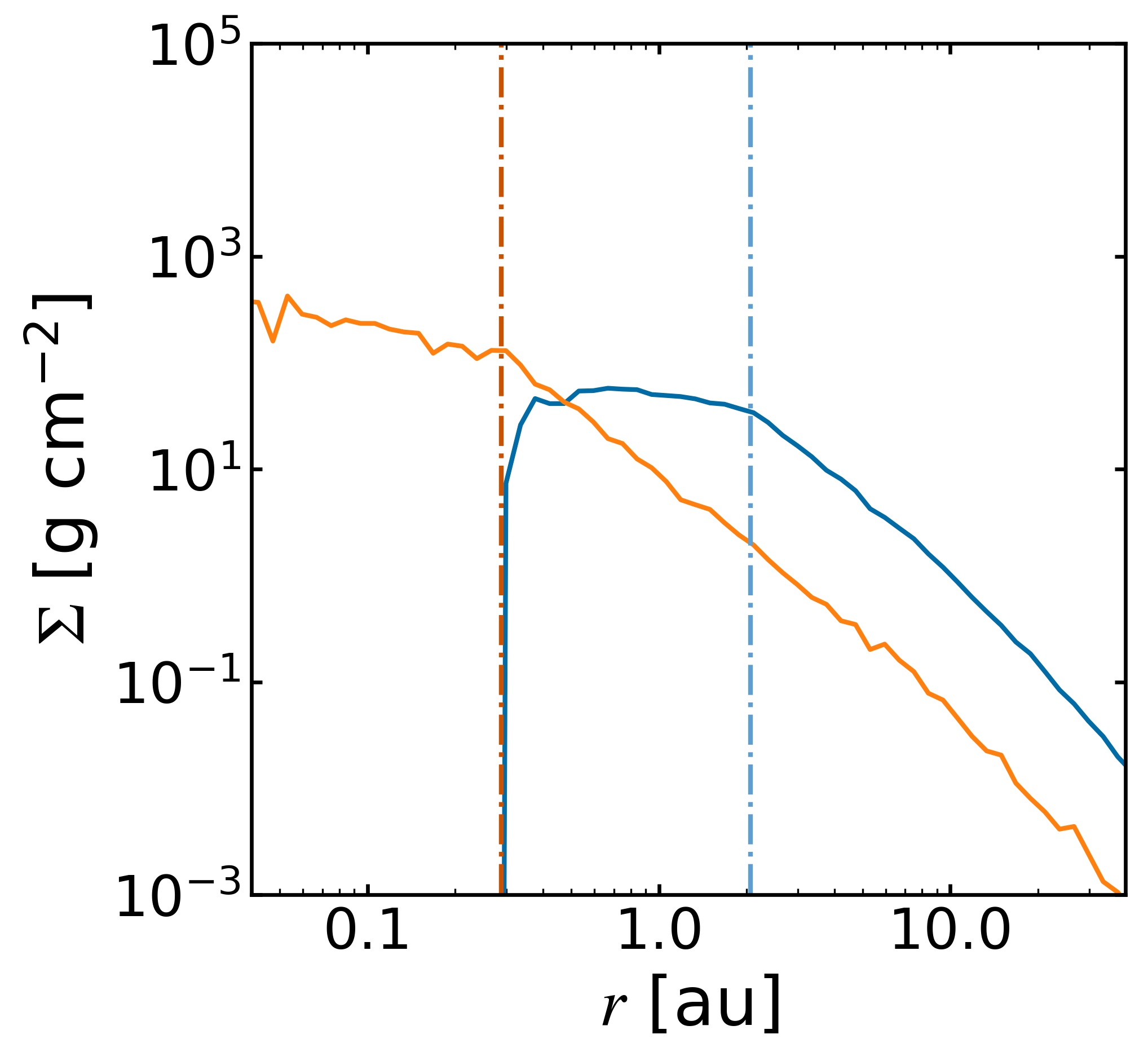}{0.25\textwidth}{(h)}
              }
    \caption{Snapshots of silicate particles in the $r$-$(z/r)$ plane and their surface densities. 
The upper panels show the snapshots 
of amorphous (the blue dots) and crystalline (the orange dots) silicates
at (a) \(t = 1.8\times 10^4\) years (\(\sim 0.5\, t_{\rm diff,snow}\)), (b) \(t=3.6\times 10^4\) years (\(\sim 1.0 \, t_{\rm diff,snow}\)), (c) \(1.8 \times 10^5\) years (\(\sim 5 \, t_{\rm diff,snow}\)) and (d) \(9.0 \times 10^5\) years (\(\sim 25 \, t_{\rm diff,snow}\)). 
The snow and annealing lines are at 2.0 and 0.28 au, which are
    represented by the light blue and red dash-dotted lines. 
 The lower panels show the surface densities of
 amorphous (blue solid line) and crystalline (orange solid line) silicates
 corresponding to the upper panels.
    }
\label{fig:snap}
\end{figure*}

We first show the results 
in a simple case without the sticking to pebbles and
with time-constant $F_{\rm p/g}$. 
Figure~\ref{fig:snap} show snapshots 
of amorphous (the blue dots) and crystalline (the orange dots) silicate particles
and their surface density evolution
in a run with fiducial parameters, $\alpha=10^{-2}$,
$\dot{M}_{\textrm{g}}=10^{-7}\, M_{\odot}\,\textrm{yr}^{-1}$, and $F_{\rm p/g}=0.1$.
We always adopt $M_* = M_\odot$, $L_* = L_\odot$, and $f_{\rm sil}=0.5$,
in this paper.
With these parameters,
the disk is viscous-heating dominated at $r < 7.1$ au
and irradiation dominated at $r > 7.1$ au.
The disk gas surface density (subsection~\ref{subsec:disk}) is
\begin{equation}
\Sigma_{\rm g} = \left\{
\begin{array}{ll}
1.2 \times 10^3 (r/1\,{\rm au})^{-3/5} \;\textrm{gcm}^{-2} & (r < 7.1 \,{\rm au}) \\  
3.0 \times 10^3 (r/1\,{\rm au})^{-15/14} \;\textrm{gcm}^{-2} & (r > 7.1 \,{\rm au}).
\end{array}
\right.
\label{eq:Sig_g_snap}
\end{equation}  
The snow and annealing lines are at $r=2.0$ and 0.28 au. 
We use the injection time interval $\delta t_{\rm inj} = 1.0$ year 
in all the runs in this paper
(in the runs with decaying pebble flux, $\delta t_{\rm inj}$
is increased after the decay starts). 
The corresponding super-particle mass is $m = 1.0\times 10^{25}\, {\rm g}$. 

Figure~\ref{fig:snap}a-d are the snapshots
at $t =$ \(1.8\times 10^4\), \(3.6\times 10^4\), \(1.8\times 10^5\), and \(9.0\times 10^5\) years.
The characteristic radial diffusion timescale is
$t_{\rm diff} \sim r^2/(6 \nu {\cal R}^2) =
r^2/2 \nu \sim (2 \alpha h_{\rm g}^2)^{-1} \Omega^{-1}$
(Eqs.~\ref{eq:diffx} and \ref{eq:diffy}).
In the unit of $t_{\rm diff,snow}$, the diffusion timescale at $r=r_{\rm snow}$,
the time in panels a-d correspond to
$t \sim 0.5 \, t_{\rm diff,snow}$, $1.0 \, t_{\rm diff,snow}$,
$5 \,  t_{\rm diff,snow}$, and $25 \, t_{\rm diff,snow}$, respectively.
Silicate particles released at the snow line
diffuse both inward and outward with a net inward mean flow inside the snow line. 
Accordingly, the surface density distribution of silicate particles,
which is calculated from the snapshot with Eq.~(\ref{Svis}),
increases with time.
We find that the surface density distribution  
at $t = 25 \, t_{\rm diff,snow}$ (panel d) does not change any more 
even if simulations are continued, implying that
the distribution at $t = 25 \, t_{\rm diff,snow}$ is already in equilibrium 
in the range of $r$ in the plot ($r < 40$ au). 

At $t \sim 0.5 \, t_{\rm diff,snow}$ (panel a), 
about a half of silicate particles released at $t\sim 0$ arrive at the annealing line
and the crystalline silicate surface density increases
to a half of that in the equilibrium state at $r \sim r_{\rm annl}$.
This timescale is explained as follows.
The radial drift timescale for silicate particles 
by disk gas accretion flow is $\sim r/v_r \sim 2 r^2/3 \nu \sim t_{\rm diff}$.
In the viscous heating dominant region, 
$h_{\rm g}\propto r^{-1/20}$ and accordingly $t_{\rm diff} \propto 
h_{\rm g}^{-2} \Omega^{-1} \propto r^{7/5}$;
the diffusion spends more time at larger $r$.
Thereby, the total timescale for particles released at $r\sim r_{\rm snow}$
to migrate to the annealing line would be $\sim t_{\rm diff,snow}$,
which agrees with the simulation result within a factor of 2.
At $t \sim 1.0 \, t_{\rm diff,snow}$  (panel (b)), roughly 5 \% of crystalline silicate particles come back to the snow line
and the crystalline silicate surface density increases
to a half of that in the equilibrium state at $r \sim r_{\rm snow}$.
The timescale to diffuse out from $r \sim r_{\rm annl}$
to $r \sim r_{\rm snow}$ is also $\sim t_{\rm diff,snow}$,
which is consistent with the simulation result in panel b.
Although, in fact, the diffusion timescale for inward transport would be shorter than outward transport by inward disk accretion, accretion timescale may be comparable to diffusion timescale for transport from the snow line and the annealing line and the simple estimation with diffusion timescale helps to roughly interpret the simulation results.
At $t \sim 5 \, t_{\rm diff,snow}$ (panel c), 
the crystalline silicate surface density reaches the equilibrium
value at $r \sim r_{\rm snow}$, but it has not 
reached the equilibrium value in outer region, because
$t_{\rm diff}$ is longer for larger $r$.

\begin{figure*}
    \gridline{\fig{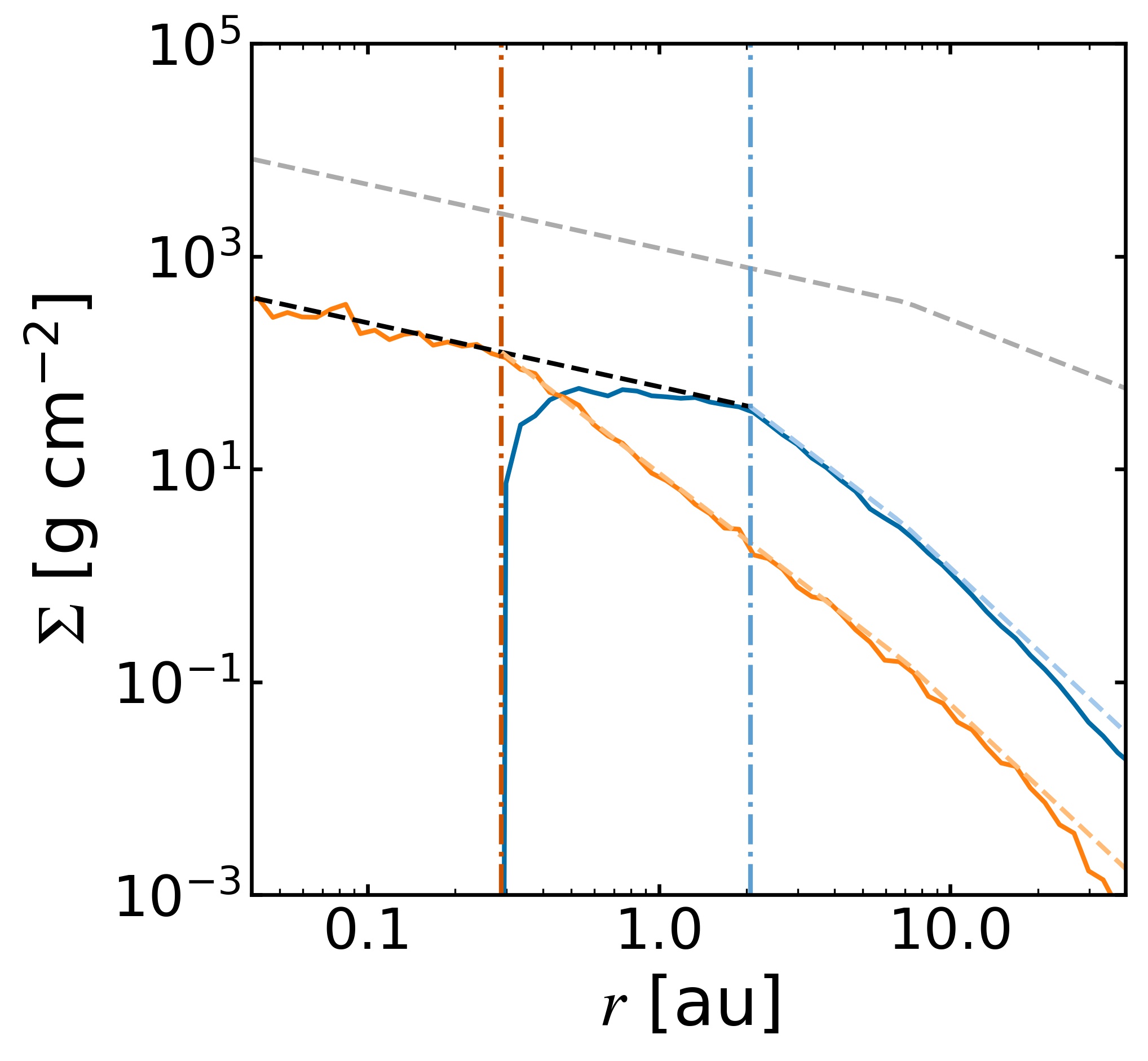}{0.45\textwidth}{(a)}
              \fig{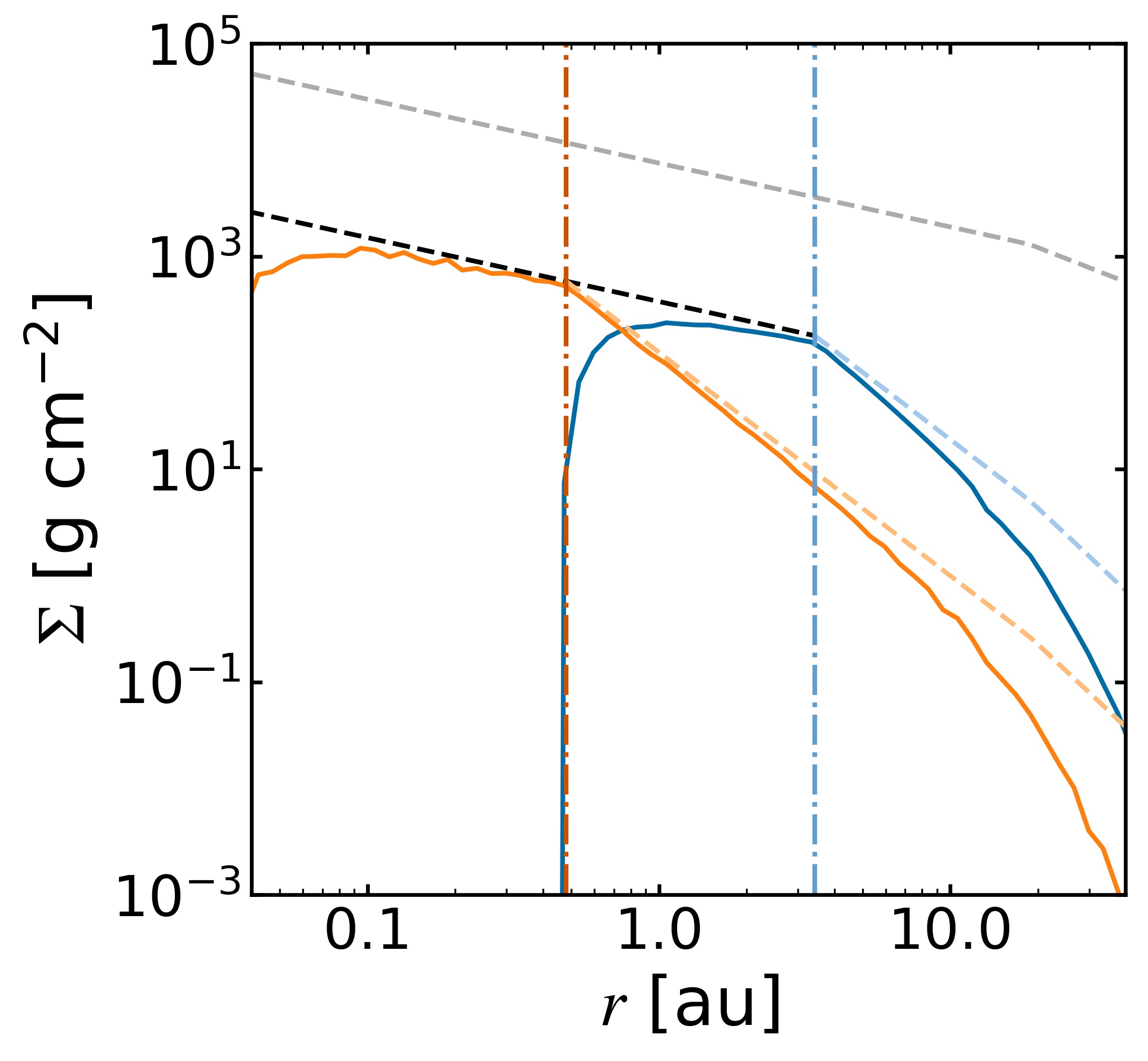}{0.45\textwidth}{(b)}
             }
    \caption{The surface densities of silicates without sticking to pebbles (a) for \(\alpha=10^{-2}\) and \(\dot{M}_{\textrm{g}}=10^{-7}\, M_{\odot}\,\textrm{yr}^{-1}\) and (b) for \(\alpha=10^{-3}\) and \(\dot{M}_{\textrm{g}}=10^{-7}\, M_{\odot}\,\textrm{yr}^{-1}\) after \(10^6\) years. The orange and blue solid line indicate the surface density of crystalline silicates and amorphous silicates. The gray dashed line indicates the gas surface density given by Eq.~(\ref{eq:Sig_g_snap}). The black dashed line inside the snow line is the analytical steady accretion solution given by Eq.~(\ref{eq:Sig_sil_steady}). The orange and light blue dashed lines are the analytical zero net flux solutions (Eq.~(\ref{eq:zero_net_index})) departing from the values of the steady accretion solution at the annealing and snow lines, respectively. The red and light blue dash dotted line 
represent $r_{\rm annl}$ and $r_{\rm snow}$. We calculated surface densities 
from the particle distributions obtained by the Monte Carlo simulations,
using Eq.~(\ref{eq:sigcal}). \label{fig:nopeb}}
\end{figure*}

Figure \ref{fig:nopeb}a shows the silicate particle surface density
at $t = 1.0\times 10^6$ years of the run in Figure~\ref{fig:snap},
where the equilibrium state is established. 
The orange and blue lines are the surface densities
of crystalline and amorphous silicate particles 
calculated by Eq.~(\ref{eq:sigcal}) from
the particle distributions obtained by the Monte Carlo simulations,
In this plot, the blue dashed line represents 
the assumed gas surface density given by Eq.~(\ref{eq:Sig_g_snap}).
The other dashed lines represent
the analytically predicted surface densities of silicate particles, as explained below.
These analytical predictions reproduce
the simulation results very well.
The magenta and olive dashed lines indicate the analytical zero net flux solutions
that depart from the values of the steady accretion solution, 
given by Eq.~(\ref{eq:Sig_sil_steady}),
at the annealing and snow lines, respectively. 
The radial power-law index of the zero net flux solution ($\Sigma_{\rm sil} \propto r^{-q}$) is
given by $q = p + 3/2$ 
(Eq.~(\ref{eq:zero_net_index})) where $p =3/5$ and $q = 21/10$
in the viscous-heating dominated regime
and $p =15/14$ and $q = 18/7$
in the irradiation dominated regime (Eq.~\ref{eq:Sig_g_snap}).
The sum of numerically obtained silicate surface densities
of the crystalline (the orange line) and amorphous (the blue line) 
agrees with the analytical steady accretion solution inside the snow line.
Taking account of the decrease due to the transformation from amorphous to crystalline silicates at $r \sim r_{\rm annl}$, the simulation result shows that
the amorphous and crystalline silicate surface densities
individually follow the steady accretion solution 
inside the snow and annealing lines, respectively.
On the other hand, they fit with the zero net flux solution
outside the snow and annealing lines, respectively.
In Figure \ref{fig:nopeb}b, we adopt $\alpha = 10^{-3}$, so that
$t_{\rm diff}$ is 10 times longer and 
$\Sigma_{\rm sil}$ has not reached the equilibrium state beyond the snow line.
Therefore, the equilibrium surface densities of crystalline silicates \(\Sigma_{\textrm{cry,sil}}\) and the sum of crystalline and amorphous silicates
\(\Sigma_{\textrm{tot,sil}}\equiv\Sigma_{\textrm{cry,sil}}+\Sigma_{\textrm{amr,sil}}\) obtained by the Monte Carlo simulations are fitted as
\begin{equation}
    \label{Sigmacry}
    \Sigma_{\textrm{cry,sil}}(r) =\left\{
        \begin{array}{ll}
            \Sigma_{\rm sil,0}(r) & [\textrm{for } r\le r_{\rm annl}]\\
             \displaystyle \Sigma_{\rm sil,0}(r)\left(\frac{r}{r_{\textrm{annl}}}\right)^{-3/2}& [\textrm{for } r>r_{\rm annl}],
        \end{array} 
    \right.
\end{equation}
and
\begin{equation}
    \label{Sigmaamr}
    \Sigma_{\textrm{tot,sil}}(r) =\left\{
        \begin{array}{ll}
            \Sigma_{\rm sil,0}(r) & [\textrm{for } r\le r_{\textrm{snow}}]\\
            \displaystyle \Sigma_{\rm sil,0}(r) \left(\frac{r}{r_{\textrm{snow}}}\right)^{-3/2}& [\textrm{for } r>r_{\textrm{snow}}],
        \end{array} 
    \right.
\end{equation}
where 
\begin{equation}
\Sigma_{\rm sil,0}(r) \equiv f_{\textrm{sil}}F_{\textrm{p/g}}\Sigma_{\textrm{g}}(r)
= 0.05 \left(\frac{f_{\textrm{sil}}}{0.5}\right)
          \left(\frac{F_{\rm p/g}}{0.1}\right) \Sigma_{\textrm{g}}(r).
\end{equation}
At $r > r_{\rm snow}$, 
because $\Sigma_{\rm cry,sil} \ll \Sigma_{\rm amr,sil}$, 
\begin{equation}
\Sigma_{\rm amr,sil} (r) \simeq 
\Sigma_{\rm sil,0}(r) \left(\frac{r}{r_{\textrm{snow}}}\right)^{-3/2}
\hspace{3mm}  [\textrm{for } r>r_{\textrm{snow}}].
 \label{eq:Sigmaamr2}
\end{equation}

These surface density distributions are physically interpreted as follows.
We release amorphous silicate particles at the snow line.
While the outward diffusion length increases with the square root of $t$,
inward drift length due to disk gas accretion increases with $t$.
Eventually, they balance to establish an equilibrium zero net flux distribution.
After that, the total mass flux of
the particles injected at different times that are passing the inner edge 
becomes equal to the given injection mass flux at the snow line ($\dot{M}_{\rm sil} = f_{\rm sil} F_{\rm p/g} \dot{M}_{\rm g}$). 
Thereby, amorphous silicate surface density follows
the zero net flux solution outside the snow line and
the steady accretion solution inside it.
At the annealing line, crystalline silicate particles are generated
from the drifting amorphous silicate particles with mass flux 
$\dot{M}_{\rm sil}$, which is similar to the injection of silicate particles
at the snow line due to ice sublimation.
Accordingly, crystalline silicate surface density follows
the zero net flux solution outside the annealing line and
the steady accretion solution inside it.

Equations~(\ref{Sigmacry}) and (\ref{Sigmaamr})
show that for $r>r_{\rm snow}$, independent of $r$,
\begin{align}
\frac{\Sigma_{\rm cry,sil}(r)}{\Sigma_{\rm tot,sil}(r)} \simeq 
\left( \frac{r_{\rm annl}}{r_{\textrm{snow}}}\right)^{3/2}.
\label{eq:crt_amor}
\end{align}
When the snow line is within the viscous heating dominated region, \(r_{\textrm{annl}}/r_{\textrm{snow}}\simeq 0.14\).
Equation~(\ref{eq:crt_amor}) shows $\Sigma_{\rm cry,sil}/\Sigma_{\rm tot,sil} \simeq 0.05$.
We note that the value of
$(r_{\textrm{annl}}/r_{\textrm{snow}})$ depends only on radial gradient of $T$.
It is independent of other detailed disk parameters, such as \(\alpha\) and \(\dot{M}_{\textrm{g}}\), and pebble parameters, such as $f_{\rm sil}$, $F_{\rm p/g}$,
and $\tau_{\rm s,sil}$ as long as $\tau_{\rm s,sil} < \alpha$.
This robust, predicted value of the crystalline abundance in silicates in this simple case,
 $\Sigma_{\rm cry,sil}/\Sigma_{\rm tot,sil} \simeq 0.05$, is
 smaller than the observationally inferred abundance,
 $\sim 0.3$ in \cite{Sitko2011} and $\sim 0.1$--0.6 in \cite{Sninnaka2018}. 
 It strongly suggests that 
the effects of sticking to drifting pebbles
 and time-dependent $F_{\rm p/g}$ must be incorporated into
 the simulations to predict the crystalline abundance in comets. 
  
\subsection{Case with sticking to icy pebbles} \label{subsec:peb}

\begin{figure*}
    \gridline{\fig{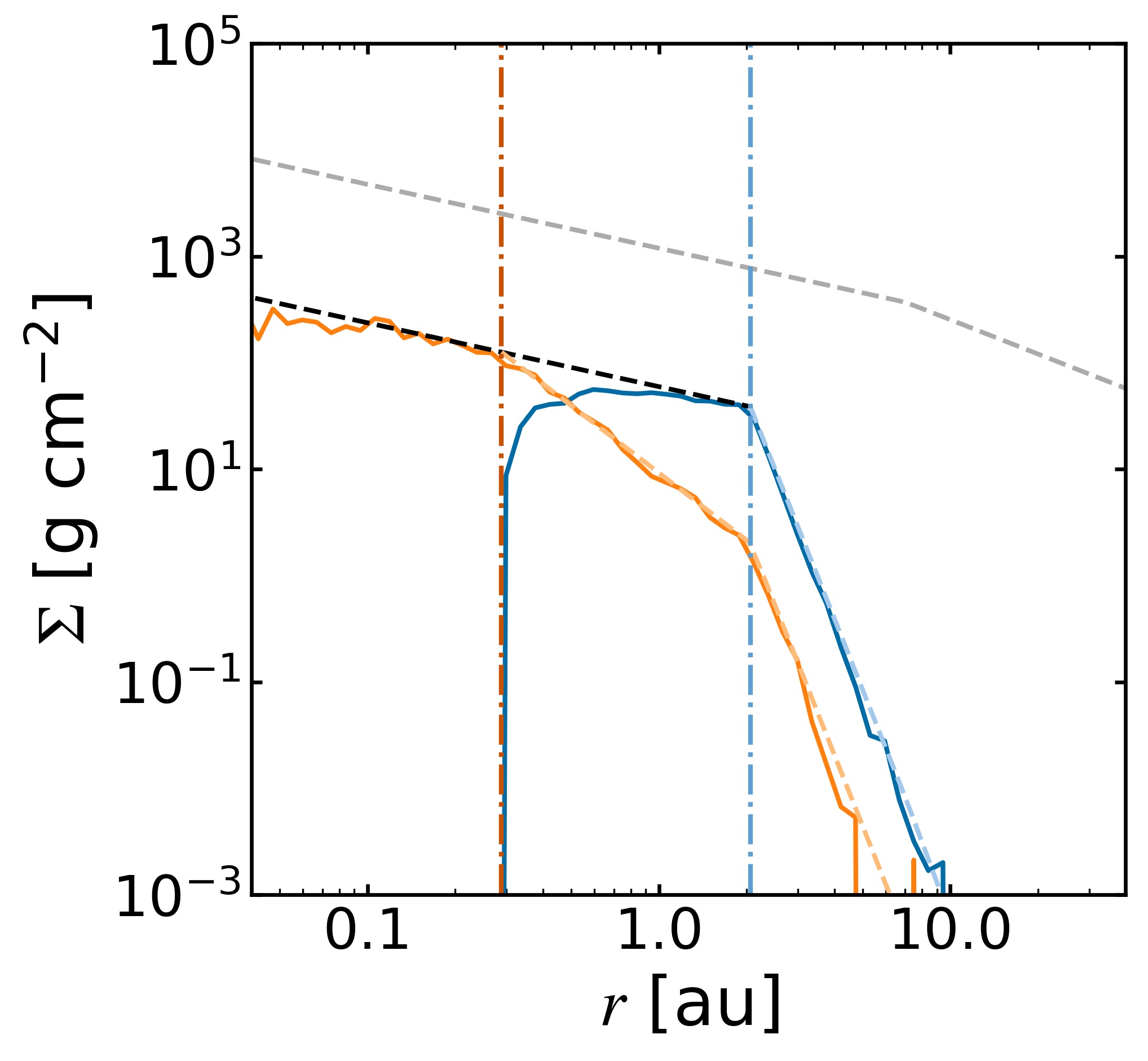}{0.45\textwidth}{(a)}
              \fig{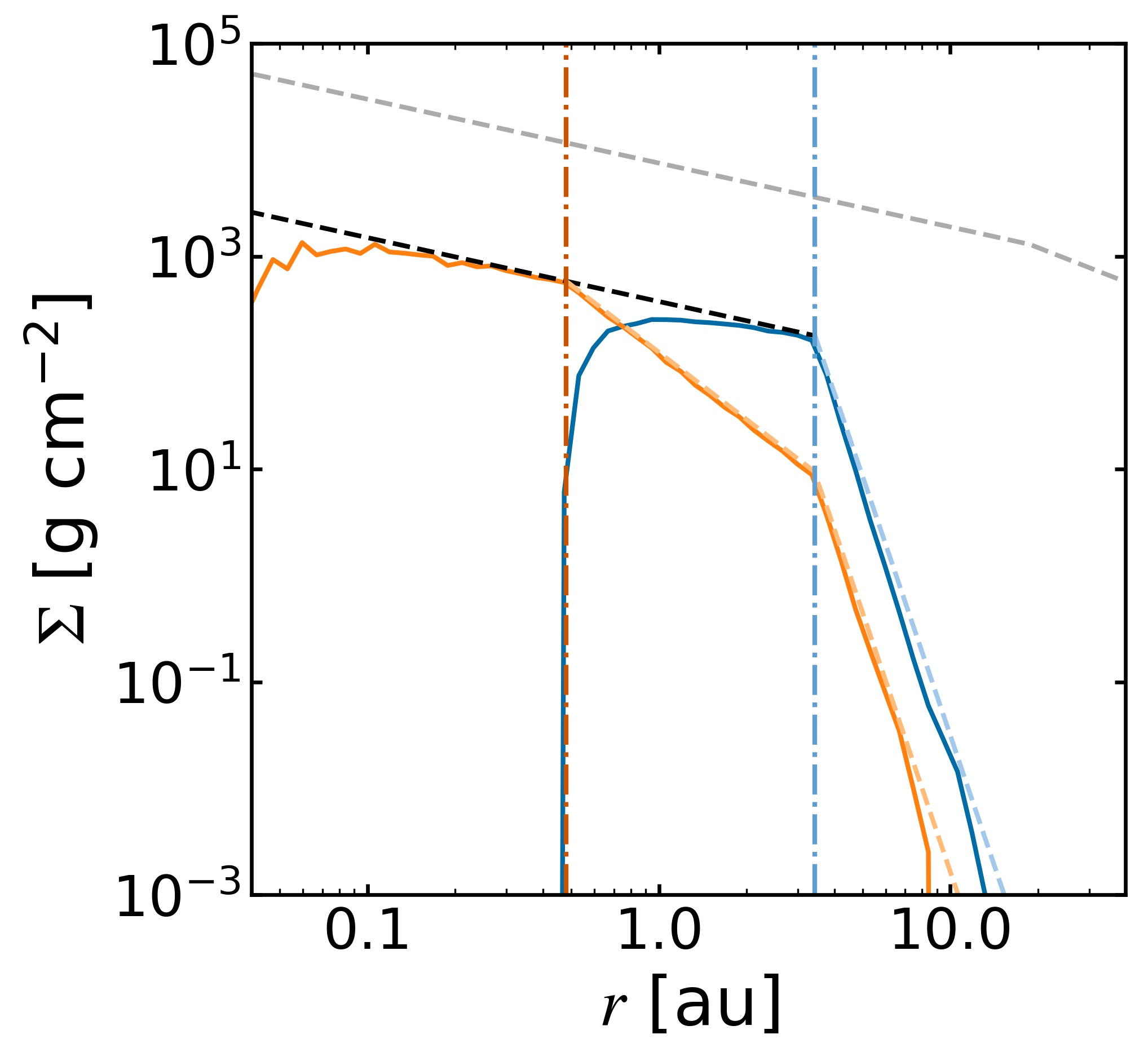}{0.45\textwidth}{(b)}
             }
    \caption{The surface densities of silicates with sticking to pebbles.
The parameters for (a) and (b)
and the meanings of the lines are the same as Fig.~\ref{fig:nopeb}
}
\label{fig:peb1}
\end{figure*}

Figure \ref{fig:peb1} shows 
the results with the effect of sticking to pebbles
after the establishment of an equilibrium state.
Comparing to the simple case without the sticking effect (section \ref{subsec:nopeb}), 
the decay of the silicate surface densities with $r$ is faster
outside the snow line, by the effect of rapid inward drift with $\tau_{\rm s,peb}$
after the sticking. 
The radial gradient of the silicate surface densities outside the snow line
are analytically predicted by Eq.~(\ref{eq:Sigwipeb})
as $q \simeq 6.4$ for $\alpha = 10^{-2}$ and $q \simeq 7.7$ for $\alpha = 10^{-3}$
in the viscous-heating dominated regime.
In the irradiation dominated regime, $q$ is larger by $15/14 - 3/5 \simeq 0.47$.
The numerical results agree with the analytical prediction.  
Silicate particles have a steeper radial gradient
than that without the effect of the sticking to pebbles
(the power-law index is larger by $\sim 4-6$).
However, because we assume that the sticking probability is the same for crystalline and amorphous silicates, their surface density distributions are smaller 
by the same factor as in the non-sticking case, and 
$\Sigma_{\rm cry,sil}/\Sigma_{\rm tot,sil}$ is still given by Eq.~(\ref{eq:crt_amor})
as $\sim 0.05$.

So far, we have considered a time-independent pebble mass flux, equivalently,
a time-independent sticking probability.
However, the pebble flux should start decaying
after the pebble formation front reaches the characteristic disk size
and the solid materials in the disk are reduced. 
Because the sticking effect substantially modifies the silicate surface density
outside the snow line, the time dependence of the pebble flux changes
the $\Sigma_{\rm cry,sil}/\Sigma_{\rm tot,sil}$ ratio, which we investigate below.

\subsection{Case with decaying pebble flux} \label{subsec:decay}

\begin{figure*}
    \gridline{\fig{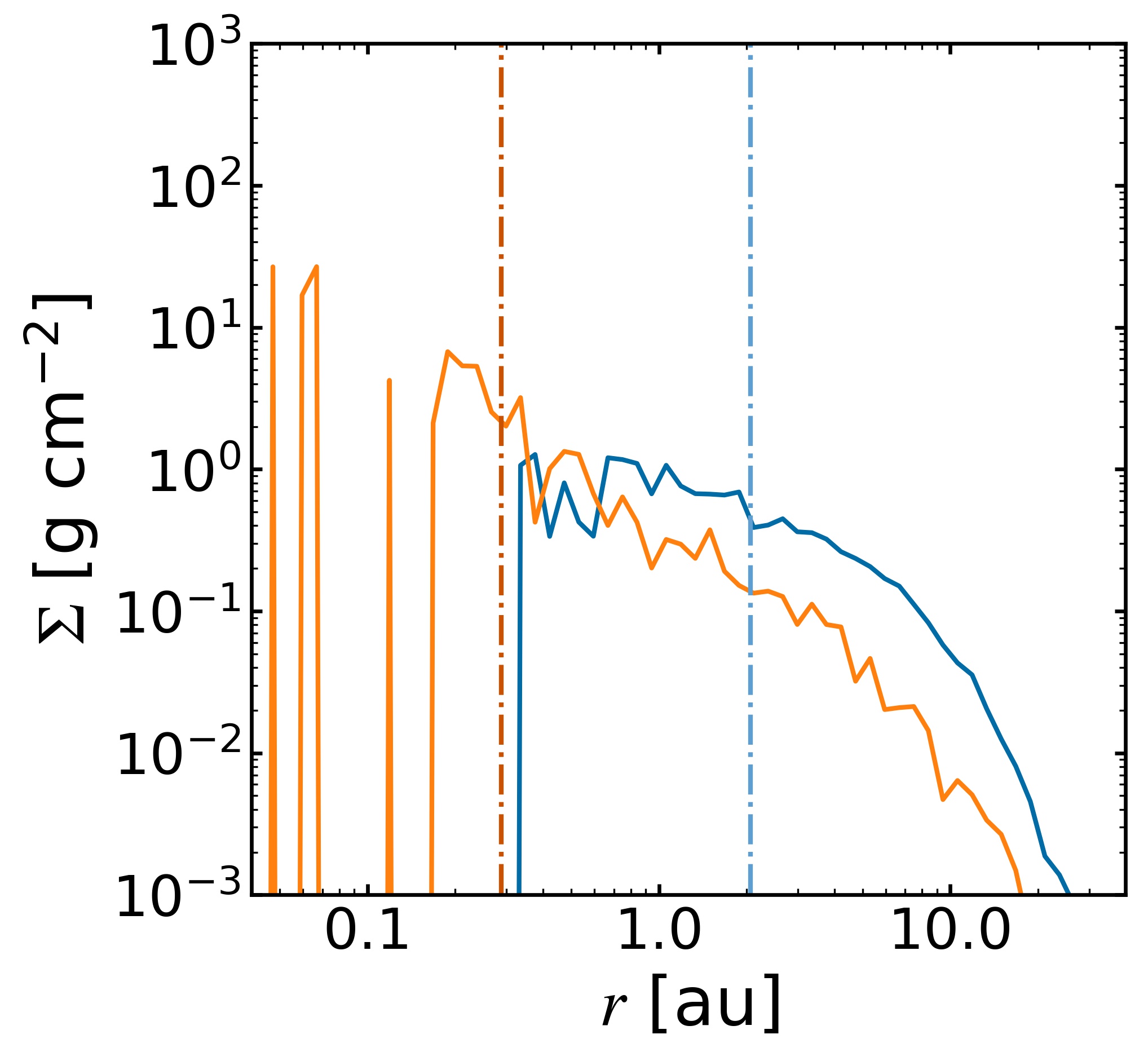}{0.3\textwidth}{(a)}
              \fig{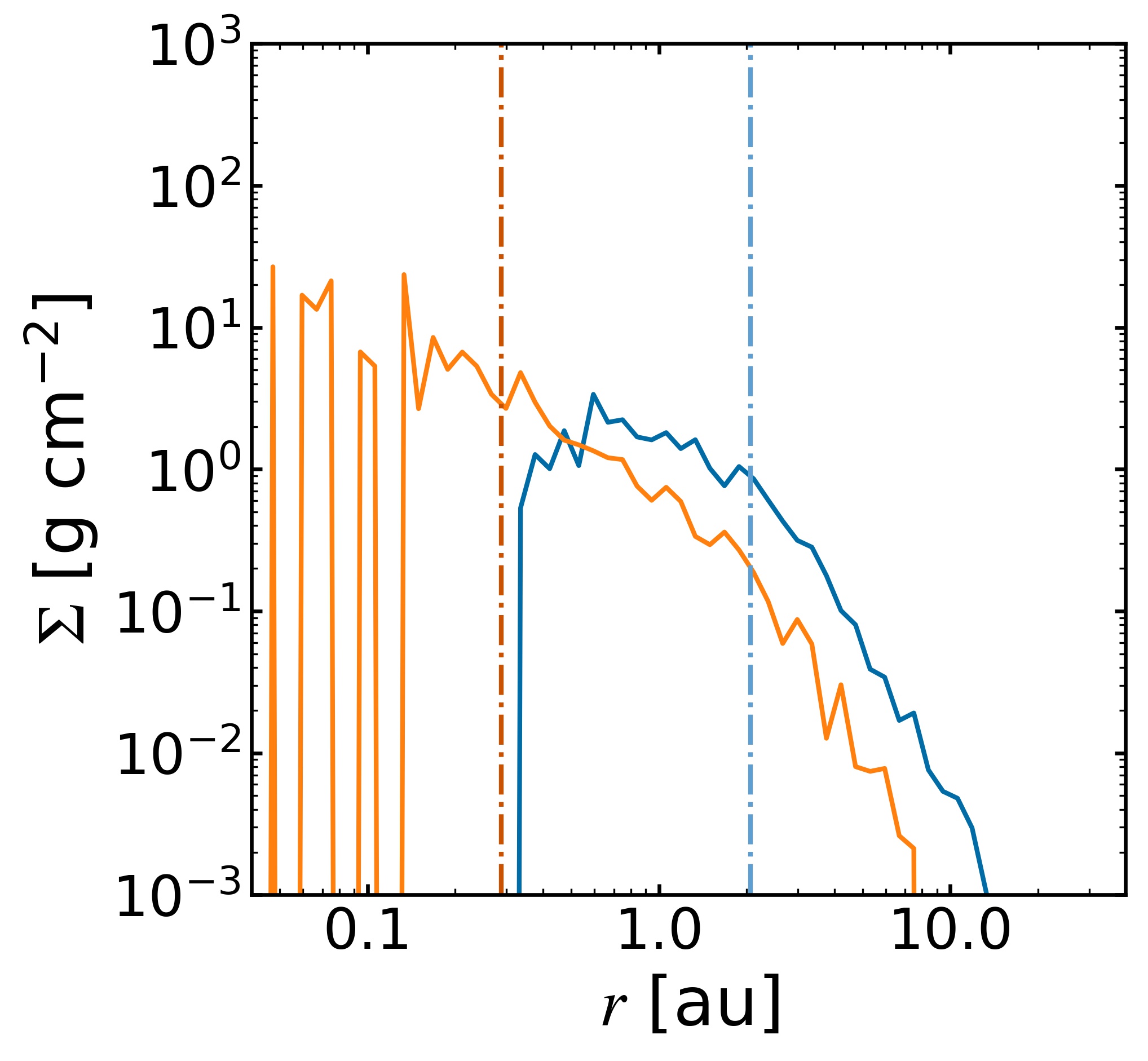}{0.3\textwidth}{(b)}
              \fig{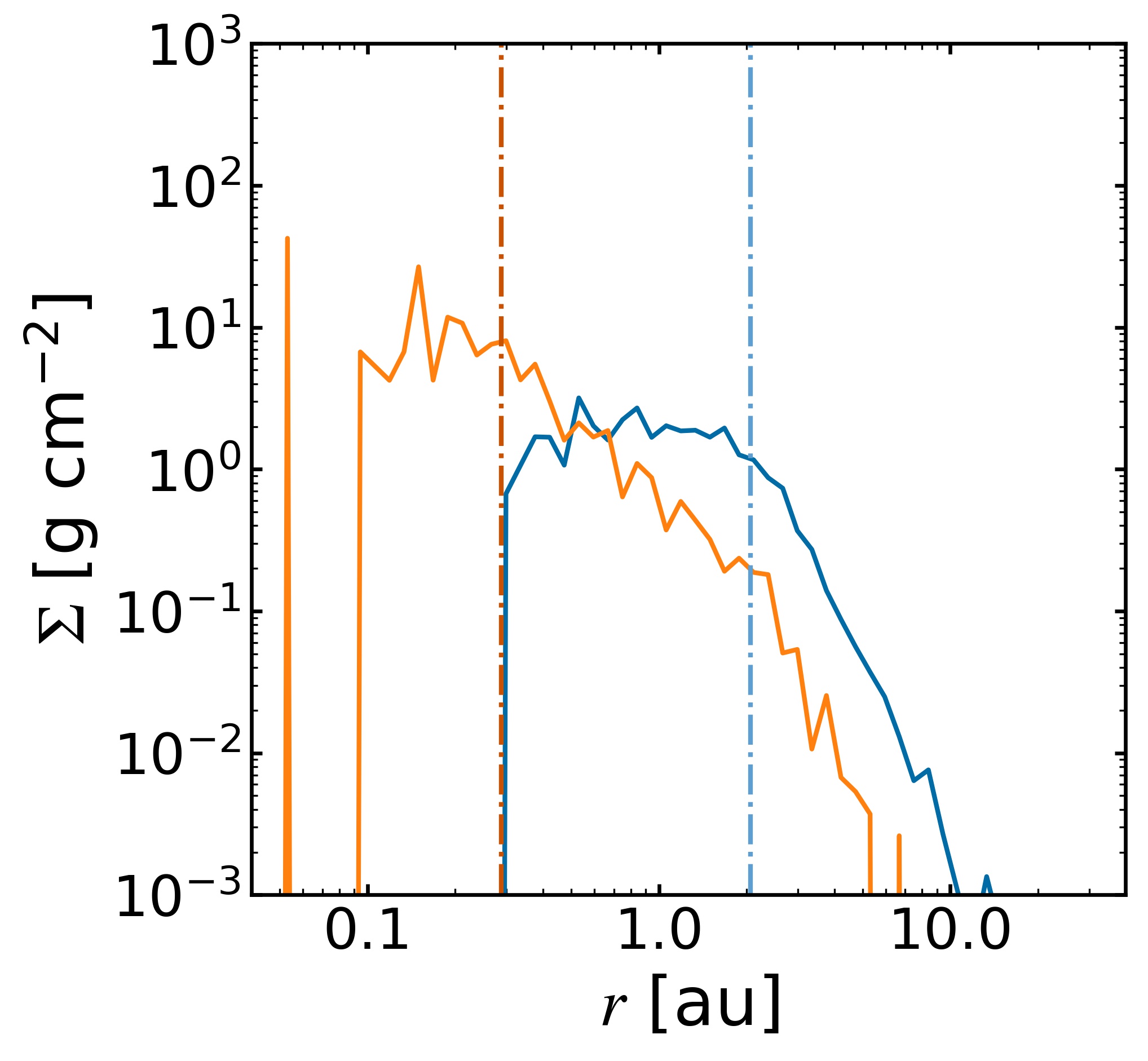}{0.3\textwidth}{(c)}
             }
    \gridline{\fig{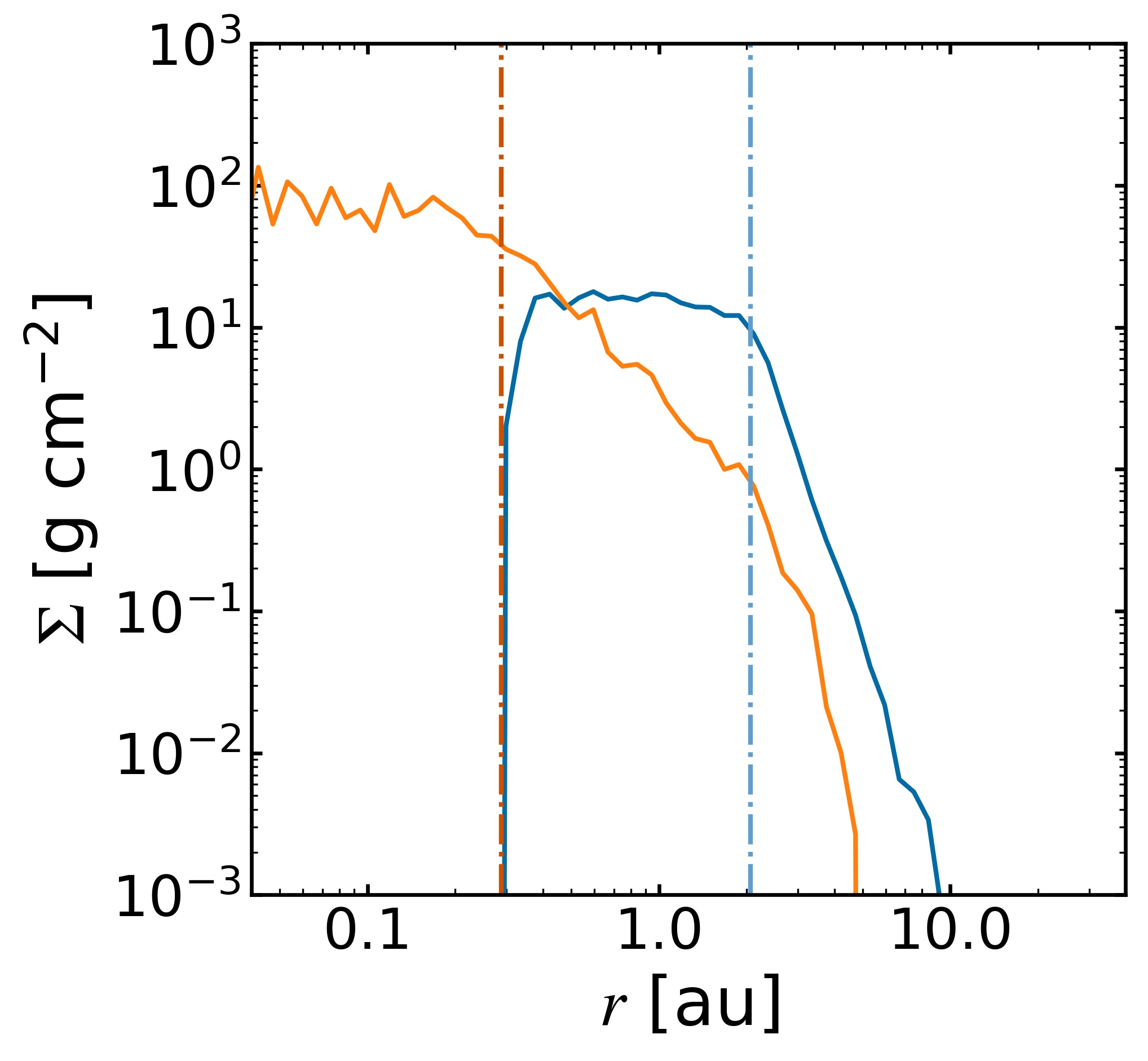}{0.3\textwidth}{(d)}
              \fig{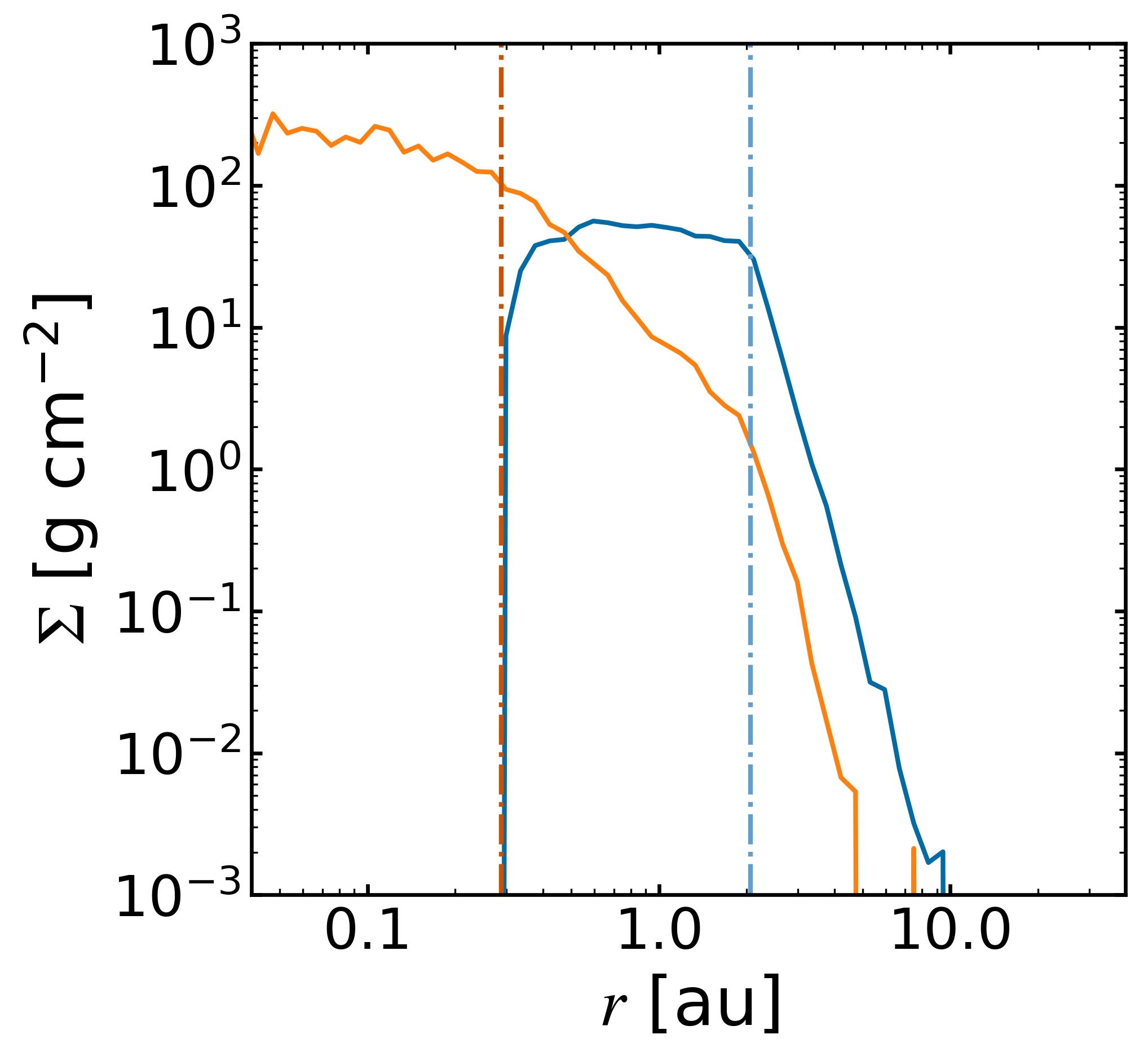}{0.3\textwidth}{(e)}
             }
    \caption{The surface density of silicates with sticking to pebbles in pebble flux attenuation for \(\alpha=10^{-2}\), \(\dot{M}_{\textrm{g}}=10^{-7}\,M_{\odot}\,\textrm{yr}^{-1}\) and (a) \(t_{\textrm{peb}}=0\) years, (b) \(t_{\textrm{peb}}=0.5 \ t_{\rm diff,snow}\), (c) \(t_{\textrm{peb}}=t_{\rm diff,snow}\), (d) \(t_{\textrm{peb}}=2 \ t_{\rm diff,snow}\), and (e) \(t_{\rm peb}=\infty\). 
The meanings of the lines are the same as Fig.~\ref{fig:nopeb}
The abundance of crystalline silicates at the snowline are about (a) 0.25, (b) 0.2, (c) 0.13, (d) 0.1, and (e) 0.05. \label{fig:peb2}}
\end{figure*}

\cite{Sato2016} and \cite{IdaYmaOku2019} showed that 
icy pebble flux decays rapidly after the pebble formation front
reaches the characteristic disk radius ($r_{\rm disk}$) at 
$t \simeq 2.0\times 10^5(r_{\rm disk}/100\,{\rm au})^{3/2}$ years,
in the fast limit of pebble growth with perfect accretion.
According to this result,
we perform simulations in which the pebble flux starts decaying
from the steady distribution in section~\ref{subsec:peb}.
We set the decaying pebble flux as
\begin{equation}
    F_{\textrm{p/g}}=F_{\textrm{p/g,0}}\exp(-t_1/t_{\textrm{peb}}),
\end{equation}
where $t_1 = t -t_0$ is the time from which the decay is switched on at $t=t_0$
and $t_{\rm peb}$ is the decay timescale
with a typical value of a few $\times 10^5$ years 
in the case of $r_{\rm disk}\sim 30\,{\rm au}$
\citep{Sato2016, IdaYmaOku2019}.
We also perform a simulation with $t_{\rm peb}=0$ as an extreme case, which means sudden truncation of pebble flux. 
To keep the mass of individual super-particles unchanged, we inject particles with a time interval,
$\delta t_{\textrm{inj}}=\delta t_{\textrm{inj,0}}\exp(t_1/t_{\textrm{peb}})$.

As shown in section~\ref{subsec:nopeb}, 
the typical timescale for crystalline silicates 
to return to the snow line, after they are released at the snow line
and pass the annealing line, is 
\begin{equation}
    \label{eq:tdiff}
   t_{\rm diff,snow} \sim \left. \frac{r^2}{\alpha H_{\rm g}^2\Omega} \right|_{r_\textrm{snow}} = 
   \left. \frac{1}{\alpha h_{\rm g}^2 \Omega} \right|_{r_\textrm{snow}}  \sim 3.6 \times 10^4 \left( \frac{\alpha}{10^{-2}} \right)^{-10/9} \left( \frac{\dot{M}_{\textrm{g}}}{10^{-7}\,M_{\odot}\,\textrm{yr}^{-1}}\right)^{2/9} \textrm{years}.
\end{equation}
Most of the amorphous silicates beyond the snow line may 
directly diffuse out from the snow line. 
It is expected that crystalline silicates beyond the snow line
were released earlier on average by $\sim t_{\rm diff,snow}$ than amorphous silicates there.
Because the crystalline silicate distribution would reflect an earlier pebble flux,
$\Sigma_{\rm cry,sil}/\Sigma_{\rm tot,sil}$ for decaying $F_{\textrm{p/g}}$ would be
higher than the ratio in the case of the time-independent $F_{\rm p/g}$(Eq.~(\ref{eq:crt_amor})).

According to the above prediction, we
perform simulations with various values of $t_{\rm peb}$ relative to $t_{\rm diff,snow}$,
from the equilibrium states with $F_{\rm p/g}=0.1$.
Figure \ref{fig:peb2} show the distributions of $\Sigma_{\rm cry,sil}$ and $\Sigma_{\rm amr,sil}$
at $t_1 = 2.5 \, t_{\rm diff,snow}$
of (a) $t_{\rm peb} = 0$, (b) $1.8 \times 10^4$ years ($\sim 0.5\, t_{\rm diff,snow}$), (c) $3.6 \times 10^4$ years ($\sim t_{\rm diff,snow}$), and (d) $7.2\times 10^4$ years ($\sim 2 \, t_{\rm diff,snow}$), 
for $\alpha=10^{-2}$ and $\dot{M}_{\textrm{g}}=10^{-7}\,M_{\odot}\,\textrm{yr}^{-1}$. 
We also show (e) the initial equilibrium state,
which corresponds to $t_{\rm peb} = \infty$.
The ratio $\Sigma_{\rm cry,sil}/\Sigma_{\rm amr,sil}$ is 
almost independent of $r$ at $r > r_{\rm snow}$ in all the panels, although $\Sigma_{\rm cry,sil}$ and $\Sigma_{\rm amr,sil}$ individually
decay according to the decay of the pebble mass flux. 
Because the pebbles capturing the silicate particles also decay,
both the crystalline and amorphous silicates
extend to outer regions with time.
The asymptotic value of $\Sigma_{\rm cry,sil}/\Sigma_{\rm tot,sil}$ 
is (a) 0.25, (b) 0.20, (c) 0.13, (d) 0.10, and (e) 0.05.
It increases with the decrease in $t_{\rm peb}/t_{\rm diff,snow}$, because
the pebble fluxes reflected by the crystalline and amorphous silicate distributions 
are more different for more rapid pebble flux decay with smaller $t_{\rm peb}$.  
Because the diffusion effect smooths out the contrast in the pebble fluxes,
$\Sigma_{\rm cry,sil}/\Sigma_{\rm tot,sil}$ is saturated at $\sim 0.25$
even in the limit of $t_{\rm peb} = 0$.

The observationally inferred crystalline abundance
of comets is $\sim 0.3$ in \cite{Sitko2011} and $\sim 0.1$--0.6 in \cite{Sninnaka2018} (section1). 
Figure \ref{fig:peb2} show that our simulations predict the abundance of crystalline silicates is 
$\sim 0.2$--0.25 for $t_{\rm peb} \la 0.5 \, t_{\rm diff,snow}$,
while it is $\la 0.1$ for $t_{\rm peb} \ga 2 \, t_{\rm diff,snow}$.
The relatively fast decay with $t_{\rm peb} \la 0.5 \, t_{\rm diff,snow}$ produces
the results that may be consistent with the observed abundance.
\citet{IdaYmaOku2019} showed that $t_{\rm peb}$ is a few $\times 10^5$ years
for a relatively compact disk with $r_{\rm disk}\sim 30 \ \rm au$,
which can be comparable to $\sim 0.5 \, t_{\rm diff,snow}$ for $\alpha \sim 10^{-3}$ (Eq.~(\ref{eq:tdiff})).
The consistency with the observed data is discussed more in the next section.

\vspace{1em}

\section{Conclusion and Discussion}

Crystalline silicates found in comets should have been formed in the disk inner region.
One of the possible transfer mechanisms of the crystalline silicates
to the disk outer region where cometary cores were formed 
is radial diffusion of silicate dust particles due to disk gas turbulence 
\citep{Ciesla2011}.
However, \citet{Ciesla2011} did not present a quantitative estimate of the crystalline abundance for the comparison with the observations of comets has not been done, 
because the supply mechanism of amorphous and crystalline silicate particles during 
planet formation was not specified.
\citet{Pavlyuchenkov2007} and \citet{Arakawa2021}
discussed an equilibrium radial distribution of crystalline materials.
Because they assumed that amorphous silicates have a stationary distribution 
with the uniform solid-to-gas ratio in the steady accretion disk,
they found that crystalline abundance is generally very low well beyond the snow line. 
However, the radial distribution of amorphous silicates must be 
derived by planet formation model for a qualitative estimate of crystalline abundance.

In this paper, adopting the ``pebble accretion" model, 
we have performed Monte Carlo simulations, taking into account the release of   
silicate dust particles from pebbles at the snow line at $T\sim 170 \, \rm K$, transformation of amorphous silicate particles to
crystalline particles at the annealing line at $T\sim 1000 \, \rm K$, sticking of silicate particles onto drifting pebbles beyond the snow line,
and attenuation of the pebble flux due to consumption of solid material reservoir in the disk outer region. 
For the gas disk, a steady accretion model with viscous heating and irradiation from the host star is assumed.
With this setting, we can quantitatively calculate the abundance of crystalline 
materials in all the silicate dust particles beyond the snow line.

In the simple case without sticking silicate particles onto drifting pebbles 
and with a steady pebble accretion,
we found 
through the Monte Carlo simulation and analytical argument
that the surface density of crystalline silicates scaled by the total surface density of crystalline and amorphous silicates is given by $\simeq (r_{\rm annl}/r_{\rm snow})^{3/2}$
uniformly beyond the snow line, independent
of disk parameters and silicate particle size, as long as $\alpha > \tau_{\rm s}$
(where $\alpha$ is the viscosity parameter and St is Stokes number of the silicate particles).
The validity of the condition, $\alpha > \tau_{\rm s}$, is discussed in Appendix.
When the viscous heating dominates at $r \sim r_{\rm snow}$, 
$r_{\rm annl}/r_{\rm snow}\simeq 0.14$ and the crystalline abundance is 
$\simeq 5$ \%. 
Although this robust value is substantially lower than the observationally inferred values,
we found that with a more realistic condition where 
the sticking to icy pebbles and the pebble flux decay
are included, the crystalline silicate abundance rises up to 20--25\%.

Our simulation shows that the crystalline silicate abundance outside the snow line 
is independent of distance from the central star, \(r\). 
This result is different from the results by \cite{Pavlyuchenkov2007} and \cite{Arakawa2021}, which showed the crystalline abundance decrease as \(r\) becomes larger. This difference results from the initial distribution of amorphous silicates.
Because amorphous silicates are released from sublimating pebbles at the snow line,
the radial gradient of the surface density is the same for both crystalline and amorphous silicates, in the framework of the pebble accretion scenario.
The observationally inferred abundance
has dispersion among comets of 10--60\% \citep[e.g.,][]{Sninnaka2018},
although observations would include some uncertainty.
There may be other factors that change the abundance, such as the evolution of a disk. 

We note that annealing can proceed even if \(T < 1000\,{\rm K}\) \citep[e.g.,][and references therein]{YamamotoTachibana2018, Ciesla2011},
while we assumed that amorphous silicates are annealed immediately at \(T = 1000\,{\rm K}\) for simplicity.
Whether or not they are annealed is determined by how long they experienced individual temperature.
It could increase the crystalline silicate abundance from that obtained with the simple treatment at
\(T = 1000\,{\rm K}\), as well as increase the dispersion among comets.
Because we track the particle motions including random walks due to gas turbulence,
we can evaluate the progress of annealing for individual particles,
which is left for our next paper.
 
In this paper, we assume a simple steady accretion disk, in order to
highlight the physics associated with the pebble accretion model.
In an early disk expansion phase, 
the snow and annealing lines are farther from the host star and
crystalline silicates can be dragged to outer regions by the expanding disk gas
\citep{Ciesla2011}.
However, if the dust particle diffusion timescale is shorter than the disk evolution timescale, 
the distribution of crystalline silicates would become the same as that in a steady accretion disk. 
The $\alpha$ value in the local turbulent diffusion can be much smaller than
the effective $\alpha$ value for global disk gas accretion (angular momentum transfer) \citep[e.g.,][]{Armitage2013, Hasegawa2017}.
The simulations coupled with the disk evolution are left to future study.

\cite{Ida2021} considered the disk model with two different 
effective $\alpha$ parameters, one for disk accretion ($\alpha_{\rm acc}$)  
and the other for turbulent diffusion ($\alpha_{\rm D}$), 
because it is possible that the disk accretion is regulated by disk wind angular momentum removal
rather than turbulent-diffusion angular momentum transfer.
{\bf 
They showed that small silicate dust particles released from sublimating icy pebbles 
undergo gravitational collapse into rocky planetesimals just inside the snow line,
if $\alpha_{\rm acc} \gg \alpha_{\rm D}$ and pebble mass flux is relatively high.
In the case of $\alpha_{\rm acc} \gg \alpha_{\rm D}$, 
crystalline silicate particles hardly diffuse out to regions much beyond the snow line.
Relatively active disk wind and relatively high pebble mass flux
would be realized in an early disk evolution phase,
implying that rocky planetesimal formation by this mechanism is 
preferred to occur in the early phase.
It could correspond to the formation of parent bodies of iron/stony meteorites.
A later phase, in which the disk wind decays ($\alpha_{\rm acc} \sim \alpha_{\rm D}$) and 
the pebble mass flux decreases, may correspond to the situation in Section~\ref{subsec:decay}.
In this case, the rocky planetesimal formation does not occur with \cite{Ida2021}'s mechanism.
If rocky protoplanets in the terrestrial planet region have not fully grown beyond
the asteroid parent body size, accretion of small dust particles is inefficient \citep[e.g.,][]{Guillot2014}
and the crystalline silicate particles diffuse out to regions much beyond the snow line.
If cometary cores are formed through icy planetesimal formation in this timing,
the cometary cores can include crystalline particles up to 20-25\%.
More discussions on a consistent scenario for rocky and icy planetesimal formation are
left for future work.
}

\begin{acknowledgements}
We thank Takafumi Ootsubo, Hideyo Kawakita, Aki Takigawa, and Shogo Tachibana
for fruitful and helpful discussions about observations and cosmo-chemistry.
We also thank the referee for the helpful comments.
This work was supported by JSPS Kakenhi 21H04512.
\end{acknowledgements}

\appendix
\section{Silicate dust's Stokes number and size with the fragmentation limit}

We set the silicate particle's Stokes number as $\tau_{\rm s}=10^{-5} \ll \alpha$.
Silicate particles can grow to some degree and
accordingly their Stokes number can increase,
after the release from sublimating icy pebbles.
However, the results here do not change, as long as diffusion is dominated over advection due to gas drag for silicate dust particles ($\tau_{\rm s} < \alpha$).
Here we estimate the maximum Stokes number of silicate particles,
when the fragmentation limit is applied.
As explained below,
\(\tau_{\rm s}\) is likely to be safely smaller than $\alpha$ for a steady accretion disk 
with our fiducial parameters, 
$\alpha = 10^{-2}$ and $\dot{M}_\textrm{g}=10^{-7}M_{\odot}/\textrm{yr}$.
   
The relative velocity \(v_{\rm rel}\) is regulated by turbulent diffusion for $\alpha \ga 10^{-3}$ \citep[e.g.,][]{Sato2016}.
In this case, 
        \begin{equation}
             v_{\rm rel}\simeq\sqrt{3\alpha \, \tau_{\rm s}} \, c_s.
        \end{equation}
        Because \(\tau_{\rm s}\) increases with the particle physical radius $R_{\rm d}$,
        the particle growth is limited 
        when \(v_{\rm rel}\) exceeds a threshold fragmentation velocity \(v_{\rm frag}\). 
        The maximum Stokes number for the fragmentation limit is given by
        $v_{\rm rel} \simeq v_{\rm frag}$ as
         \begin{equation}
             \tau_{\rm s,max} \simeq \frac{1}{3\alpha} \left(\frac{v_{\rm frag}}{c_s}\right)^2
              \simeq 3.3\times 10^{-5} \left(\frac{\alpha}{10^{-2}}\right)^{-1} 
              \left(\frac{v_{\rm frag}}{1\,\rm m/s}\right)^2 
              \left(\frac{T}{300\, \rm K}\right)^{-1},
        \end{equation}
        where we scale $v_{\rm frag}$ by a typical value of $1 \rm m/s$ 
        for silicate-silicate collisions \citep[e.g.,][]{Wada2013}.
        In the region near or inside the snow line that we are mainly concerned,
        viscous heating is dominant (see section 2.1).
        Substituting $T$ in the viscous heating region (Eq.~(\ref{Tvis}))
        into the above equation, we obtain 
        \begin{equation}
             \tau_{\rm s,max} \simeq 3.1 \times 10^{-5}\left(\frac{\alpha}{10^{-2}}\right)^{-4/5}\left(\frac{v_{\rm frag}}{1\,{\rm m/s}}\right)^2\left(\frac{\dot{M}_\textrm{g}}{10^{-7}M_{\odot}/\textrm{yr}}\right)^{-2/5}\left(\frac{r}{1\,\textrm{au}}\right)^{9/10}.
        \end{equation}
        For the fiducial parameters, 
        $\alpha = 10^{-2}$ and $\dot{M}_\textrm{g}=10^{-7}M_{\odot}/\textrm{yr}$, 
        in our disk model, $\tau_{\rm s,max} \la 1 \times 10^{-4}$ inside the snow line.
        Even if $\alpha = 10^{-3}$ is used, $\tau_{\rm s,max}$ is still smaller than $\alpha$. 
        Therefore, our assumption that we simply set $\tau_{\rm s} = 10^{-5}$ for
        silicate particles is justified.
        We refer to this estimate in sections 1, 2.2.1, 4.1, and 5.
        
        If $\dot{M}_\textrm{g}=10^{-8}M_{\odot}/\rm yr$ and $\alpha=10^{-3}$
         are assumed,
        $\tau_{\rm s,max} \simeq 5 \times 10^{-4} (r/1 \, \rm au)^{9/10}$.
        We note that this $\tau_{\rm s,max}$ is $\sim \alpha$ at a few au
        and our assumption of diffusion-dominance is marginal.
            
         The particle physical radius $R_{\rm d}$ corresponding to $\tau_{\rm s,max}$
         is as follows.
         Stokes number of a particle with the bulk density $\rho_{\rm bulk}$ which follows Epstein's law is given by 
        \begin{equation}
            \tau_{\rm s}=\frac{\rho_{\rm bulk}R_{\rm d}}{\rho_{\rm g}H_{\rm g}}=\frac{\sqrt{2\pi}\rho_{\rm bulk}R_{\rm d}}{\Sigma_{\rm g}},
             \label{eq:Stokes_num}
        \end{equation}
        where $\rho_{\rm g}, \Sigma_{\rm g}$, and $H_{\rm g}$ are
        the spatial and surface densities of the disk and the pressure scale height, respectively.        
        Substituting $\tau_{\rm s,max}$ into this equation, the typical particle size 
        with the fragmentation limit of $v_{\rm frag} = 1 \rm m/s$ in the
        viscous heating dominated regime is estimated as
        \begin{align}
         R_{\rm d} & \simeq \frac{\tau_{\rm s,max}\,\Sigma_{\rm g}}{\sqrt{2\pi}\rho_{\rm bulk}} \\
          & \simeq 4 \times 10^{-2} \left(\frac{\rho_{\rm bulk}}{3\,\rm g/cm^{-3}}\right)^{-1}
             \left(\frac{\alpha}{10^{-2}}\right)^{-8/5}\left(\frac{\dot{M}_\textrm{g}}{10^{-7}M_{\odot}/\textrm{yr}}\right)^{1/5}\left(\frac{r}{1\,\textrm{au}}\right)^{3/10} {\rm mm}.
             \label{eq:size}
        \end{align}
        In outer disk regions, where pebbles are formed, irradiation would be dominated.
        In this case, $\tau_{\rm s,max} \propto T^{-1} \propto r^{3/7}$ and
        $\Sigma_{\rm g} \propto r^{-15/14}$ (section 2.1).
        Accordingly, $R_{\rm d} \propto r^{3/7-15/14} = r^{-9/14}$.
        It means that silicate particles formed in outer regions to be embedded in pebbles are much smaller than
        that at the snow line.
        After their release at the snow line, they can grow up to the size given in Eq.~(\ref{eq:size}).
        The size released at a few au is $\sim 0.05\, \rm mm$ for $\alpha \simeq 10^{-2}$ and $\sim 3 \, \rm mm$ for $\alpha \simeq 10^{-3}$.

\bibliography{MC_silicate_ApJ}
\bibliographystyle{aasjournal}
\end{document}